\documentclass[aps,prc,amsfonts,preprintnumbers,superscriptaddress,showpacs,nofootinbib]{revtex4-1}
\usepackage{epsfig}
\usepackage{graphics}
\usepackage{amsmath}
\usepackage{amssymb}
\usepackage{mathtools}
\usepackage{dsfont}
\usepackage{bm}
\usepackage[colorlinks=true,linkcolor=blue,urlcolor=blue,citecolor=blue]{hyperref}

\begin{document}
\title{Is the ``RG-invariant EFT'' for few-nucleon systems cutoff independent?}

\author{A.~M.~Gasparyan}
\email[]{Email: ashot.gasparyan@rub.de}
\affiliation{Ruhr-Universit\"at Bochum, Fakult\"at f\"ur Physik und
        Astronomie, Institut f\"ur Theoretische Physik II,  D-44780
        Bochum, Germany}
      
\author{E.~Epelbaum}
\email[]{Email: evgeny.epelbaum@rub.de}
\affiliation{Ruhr-Universit\"at Bochum, Fakult\"at f\"ur Physik und
        Astronomie, Institut f\"ur Theoretische Physik II,  D-44780
        Bochum, Germany}

\begin{abstract}
  We consider nucleon-nucleon scattering using the
formulation of chiral effective field theory which is claimed to be 
renormalization group invariant. The
cornerstone of this framework
is the existence of a well-defined 
infinite-cutoff limit for the scattering amplitude at each order of
the expansion, which should not depend on a particular regulator form.
Focusing on the $^3P_0$ partial wave as a
representative example, we show that this requirement can in general
not be fulfilled beyond the leading order, in spite of the perturbative
treatment of subleading contributions to the amplitude. 
Several previous studies along these lines, including the
next-to-leading order calculation by Long and Yang [Phys. Rev. C84, 057001 (2011)] 
and a toy model example with singular long-range potentials by Long and van Kolck
[Annals Phys. 323, 1304-1323 (2008)],
are critically reviewed and scrutinized in detail.
\end{abstract}

\maketitle

\section{Introduction}
\label{Sec:Introduction}
The methods of chiral effective field theory (EFT) were first applied
to few-nucleon 
systems in the pioneering work by 
Weinberg~\cite{Weinberg:1990rz,Weinberg:1991um}. In the past decades,
chiral EFT has established itself as a standard tool in nuclear
physics, 
see Refs.~\cite{Bedaque:2002mn,Epelbaum:2008ga,Machleidt:2011zz,
Epelbaum:2012vx,Epelbaum:2019kcf,Hammer:2019poc} for reviews.
One of the key ingredients of the EFT approach in the few-nucleon 
sector is the resummation of  an infinite series of the iterations
of (at least) the leading-order (LO) two-nucleon-irreducible terms
to account for the non-perturbative nature
of the nucleon-nucleon (NN) interaction. 
Such a non-perturbative resummation requires a regularization of 
an infinite number of divergent diagrams, which is typically
realized by introducing an artificial ultraviolet regulator, a cutoff $\Lambda$.
This is particularly relevant for spin-triplet channels of NN
scattering, which probe the singular nature 
of the one-pion-exchange potential.
It is commonly agreed that physical observables cannot depend
on a particular choice of the cutoff as long as all possible terms
in the effective Lagrangian are taken into account.
However, the way to implement this requirement in concrete schemes based
on some kind of a systematically improvable perturbative expansion 
is not yet finally settled.

One pragmatic approach relies on using cutoffs of the order of the 
expected breakdown
scale $\Lambda_b$ of chiral EFT \cite{Lepage:1997cs,Epelbaum:2006pt}. 
The residual cutoff dependence of the scattering amplitude
is expected to become weaker as one increases the
EFT order of the calculation. This feature has indeed been verified numerically by explicit
calculations, see e.g.~Ref.~\cite{Epelbaum:2015pfa}. 
The finite-cutoff scheme has been pushed to high orders in the 
EFT
expansion, leading to remarkably accurate results, see Refs.~\cite{Reinert:2017usi,Entem:2017gor,Filin:2019eoe,Reinert:2020mcu,Filin:2020tcs}
for selected examples. Recently, first steps were made to formally 
justify this framework 
by explicitly demonstrating its renormalizability in the EFT sense~\cite{Gasparyan:2021edy,Gasparyan:2022sat}.

An alternative approach consists in enforcing cutoff independence
of the scattering amplitude
at each EFT order separately by taking the limit $\Lambda\to\infty$,
i.e.~by choosing the cutoff much larger than the EFT breakdown scale $\Lambda_b$.
In other words, one requires that the amplitude $T^{(n)}$, where $n$
denotes an EFT  order,
approaches a finite limit
\begin{equation}
 T^{(n)}_{\Lambda}\xrightarrow[\Lambda\to\infty]{}T^{(n)}_{\infty},\quad \quad
 \Lambda\frac{dT^{(n)}_{\Lambda}}{d\Lambda} \xrightarrow[\Lambda\to\infty]{}0.
 \label{Eq:fininte_limit}
\end{equation}
The amplitude is then claimed to satisfy renormalization group (RG) invariance (RGI).
In practical applications of this scheme, a finite cutoff value
$\Lambda\gtrsim\Lambda_b$ can be employed as long as one can show 
that the asymptotic behaviour of the amplitude is already reached
\cite{Hammer:2019poc}. More details on this approach and its
applications to few-nucleon systems can be found in recent review
articles \cite{Hammer:2019poc,vanKolck:2020llt}.

One of the motivations for such an approach is a quantum mechanical treatment of singular potentials,
i.e.~potentials that behave at the origin as $r^{-\alpha}$ with $\alpha\ge 2$, see 
\cite{Frank:1971xx} for a review. Such quantum mechanical
problems show similarity to chiral EFT, because the unregulated one-pion-exchange potential
in spin-triplet channels behaves
at short distances as $1/r^3$. It is known that one can obtain a unique
solution for the scattering amplitude
with such singular potentials
by constructing the so-called self-adjoined extension of the Hamiltonian.
The solution depends on one arbitrary parameter in each attractive partial wave
where the potential is singular, which can be fixed e.g.~by using the experimental 
value of the amplitude at some energy point.
In the language of EFT, this corresponds to introducing one contact interaction (a counter term)
in each attractive spin-triplet partial wave where the one-pion-exchange
potential is treated non-perturbatively. 
Consequently, the number of counter terms gets ``reduced'' compared to the  
perturbative analysis of Feynman diagrams involving multiple iterations of the leading-order 
potential, from which it follows that an infinite number of counter terms would be necessary 
in every spin-triplet partial wave.

This scheme has been criticized in
Refs~\cite{Epelbaum:2009sd,Epelbaum:2018zli,Epelbaum:2020maf,Epelbaum:2021sns} 
based on general arguments such as the absence of
an explicit transition between the perturbative and non-perturbative
regimes, issues with non-perturbative repulsive interactions,
appearance of spurious bound states, etc.
There are also technical complications that may prevent an application
of this scheme to few-nucleon systems, since large cutoff values
typically result in high computational cost needed for reaching
converged results.  For an extensive discussion of these and related
issues see Ref.~\cite{Tews:2022yfb}. 

Regardless of the above criticism, the  infinite-cutoff scheme seems
so far yielding reasonable results for NN
scattering~\cite{vanKolck:2020llt}, which
is our main focus here. In the context of chiral EFT, 
the discussed approach was first applied to NN scattering in
Ref.~\cite{Nogga:2005hy}, where the cutoff independence  of the
LO partial wave amplitudes was demonstrated by a direct numerical
calculation. 

It is also well established that a consistent inclusion of
next-to-leading-order (NLO) interactions within the RG invariant scheme is only possible
perturbatively, i.e.~using the distorted-wave Born approximation.
This is because of a singular behaviour of the NLO potential at short distances,
which becomes particularly problematic
when it is repulsive, see Refs.~\cite{PavonValderrama:2005wv,PavonValderrama:2005uj,Zeoli:2012bi,vanKolck:2020llt}.
A perturbative inclusion of the NLO terms in spin-triplet channels
within the infinite-cutoff scheme was considered
in Refs.~\cite{Long:2011qx,Long:2011xw}
using a regularization scheme in momentum space,  
see Refs.~\cite{Valderrama:2009ei,Valderrama:2011mv} for an analogous approach in coordinate space.
A numerical test of the cutoff independence was performed by varying 
the cutoff up to $\Lambda\simeq 5$~GeV.
The number of the NLO counter terms was determined based on the short-range 
behaviour of the LO wave function and of the NLO potential.
Another study in support of the infinite-cutoff scheme
was carried out in Ref.~\cite{Long:2007vp}, where the authors 
considered a toy model with the LO long-range potential $V_{\text{LO}}\sim 1/r^2$
and the NLO long-range potential $V_{\text{NLO}}\sim 1/r^4$. This 
choice of the long-range interactions allows one to perform a part of the analysis analytically.
An incomplete proof of the cutoff independence of the NLO amplitude was presented in Ref.~\cite{Long:2007vp}.
 
In the present work, we reconsider the infinite-cutoff scheme
for NN scattering at NLO in the EFT expansion
and critically examine the findings of
Refs.~\cite{Long:2011xw,Long:2011qx,Long:2007vp}. We demonstrate that
the oscillating nature of the LO wave function near the origin
caused by the singular (attractive) behavior of the LO potential generally prevents
one from achieving a cutoff-independent result for the subleading scattering
amplitude in the $\Lambda \to \infty$ limit. To keep our
considerations simple we focus here on the case of NN scattering in
the $^3P_0$ partial wave, which may serve as a representative example.  

Our paper is organized as follows.
In Sec.~\ref{Sec:general} we illustrate the above-mentioned issue of 
the infinite-cutoff scheme
based on general arguments and using a simplified version of the NN
two-pion-exchange potential.
In Sec.~\ref{Sec:LongYang} we examine in detail the NLO analysis by
Long and Yang \cite{Long:2011xw} and critically revise their
conclusions. The implications and generalizations of 
our results are discussed in Sec.~\ref{Sec:Generalizations}.
Next, in Sec.~\ref{Sec:LongVanKolck}, we consider the toy model
example of Ref.~\cite{Long:2007vp}. The complete renormalizability proofs of the
LO and NLO scattering amplitudes for this toy model are given in 
appendices \ref{Sec:AppenLO} and
\ref{Sec:AppenNLO}, respectively.  
The main results of our paper are
summarized in Sec.~\ref{Sec:sum}.

\section{General discussion}
\label{Sec:general}
In this section we present a general discussion of the issues
emerging in the infinite-cutoff scheme at NLO on a rather qualitative 
and not fully mathematically rigorous level.
These considerations are sufficient to illustrate the main idea of
our paper. The analytical considerations are supplemented 
by numerical calculations using the NN two-pion-exchange potential,
modified to exhibit a more regular behavior at short distances. This
modification allows us to simplify the presentation by reducing the
number of subtractions in the NLO amplitude. We focus on NN scattering
in the $^3P_0$ partial wave and analyze in detail the cutoff
dependence of the amplitude. This partial wave provides as a typical
example of the attractive spin-triplet channel, where the
one-pion-exchange potential is non-perturbative, 
and is particularly easy to analyze due to the absence of coupled channels.
The results can be generalized to other partial waves in a straightforward way.

\subsection{Formalism}
\label{Sec:formalism}
We start with describing the formalism. 
We consider the LO (NLO) potential $V^{\rm \scriptscriptstyle LO{}}$
($V^{\rm \scriptscriptstyle NLO{}}$)  in chiral EFT, 
with $V^{\rm \scriptscriptstyle LO{}}$
consisting of the one-pion-exchange potential
and the short-range part:
\begin{align}
V^{\rm \scriptscriptstyle LO{}}=V_{1\pi,\Lambda}+V^{\rm \scriptscriptstyle LO{}}_{\text{short},\Lambda},
\end{align}
where the subscript $\Lambda$ signifies that the corresponding potential is
regulated.
The short-range part $V^{\rm \scriptscriptstyle LO}_{\text{short},\Lambda}$ involves the $S$-wave contact
interactions and the counter terms necessary for the renormalization of the LO amplitude,
in particular, the leading contact terms in the spin-triplet channels where the
one-pion-exchange potential is attractive and treated
non-perturbatively. Here and in what follows, we 
denote by $\Lambda$ the set of all values of cutoffs
$\Lambda=\{\Lambda_i\}$ that a quantity $X_\Lambda$ depends upon.
The NLO potential is given by the
leading two-pion-exchange potential to be specified below and the short-range part
\begin{align}
V^{\rm \scriptscriptstyle NLO{}}=V_{2\pi,\Lambda}+V^{\rm \scriptscriptstyle NLO{}}_{\text{short},\Lambda}.
\end{align}
The short-range potentials consist of contact interactions multiplied by
the corresponding low-energy constants (LECs),
\begin{align}
V^{\rm \scriptscriptstyle LO}_{\text{short},\Lambda}&=C_0 V^{\rm \scriptscriptstyle LO}_{\text{ct},0,\Lambda}+\dots,\nonumber\\
V^{\rm \scriptscriptstyle NLO{}}_{\text{short},\Lambda}&=\sum_{i=0}^n C^{\rm \scriptscriptstyle NLO{}}_{2i} V^{\rm \scriptscriptstyle NLO{}}_{\text{ct},2i,\Lambda}+\dots,
\end{align}
where we only specify explicitly terms contributing to the $^3P_0$
partial wave and omit the channel indices. The second subscript of 
$V^{\rm \scriptscriptstyle LO}_{\text{ct},0,\Lambda}$ and $V^{\rm
  \scriptscriptstyle NLO{}}_{\text{ct},2i,\Lambda}$ gives the power of
momenta counted relative to the LO contribution, i.e.~$V^{\rm
  \scriptscriptstyle LO}_{\text{ct},0,\Lambda} 
\sim\vec{p} \,^2$ and 
$V^{\rm
  \scriptscriptstyle NLO{}}_{\text{ct},2i,\Lambda} \sim \vec{p} \,^{2i+2}$.
The upper limit $n$ in the sum over $i$ is determined by the number of counter terms necessary for the renormalization of the
NLO amplitude. The explicit form of the short-range potentials will be
given below. 

Some statements on the regularization are in order here.
Since the EFT formulation considered here aims at a complete
elimination of regulator dependence~\cite{vanKolck:2020llt}, the infinite-cutoff limit of the
scattering amplitude
should not depend on a particular regularization prescription.
As stated in Ref.~\cite{Song:2016ale},
``RGI requires not only independence of observables on the
numerical value of the cutoff $\Lambda$ but also independence on the
form of the regulator function itself''.
Since the potential involves several different structures, one  
always has a freedom to employ different types of the regulators and
cutoff values for each of them.

The regularized potential in the plain-wave basis is obtained from the unregularized one
by multiplying it with the corresponding form factor $F_\Lambda(\vec
p\,',\vec p \, )$.
Throughout this work, we will mostly use non-local regulators
\begin{align}
 F_\Lambda(\vec p\,',\vec p \, )=F_\Lambda(p')F_\Lambda(p),
 \label{Eq:nonlocal_regulator}
\end{align}
where $F_\Lambda(p)$ is either the smooth power-like form factor,
\begin{align}
F_\Lambda(p)&=\left( \frac{\Lambda^2}{\Lambda^2+p^2} \right)^n,
 \label{Eq:smooth_regulator}
\end{align}
with $n$ chosen sufficiently large to remove all divergences,
or the sharp cutoff,
\begin{align}
F_\Lambda(\vec p\,',\vec p \, )=\theta(\Lambda-p')\theta(\Lambda-p).
\label{Eq:sharp_cutoff}
\end{align}
The non-local regulators can be applied directly to the partial-wave
projected potentials. 
In Sec.~\ref{Sec:local}, we will also discuss the case of a locally 
regularized potential with
 \begin{align}
 F_\Lambda(\vec p\,',\vec p \, )&=\Phi_\Lambda(q^2),
 \label{Eq:local_regulator}
 \end{align}
where $\vec q=\vec p\,'-\vec p$ and $\Phi(q^2)$ is some 
smooth regulator, e.g.,
 \begin{align}
\Phi_\Lambda(q^2)&=\left( \frac{\Lambda^2}{\Lambda^2+q^2} \right)^n.
 \label{Eq:local_regulator2}
 \end{align}
It will become obvious from our analysis that 
other forms and types of regulators used in the literature
including Gaussian regulators,
spectral function regularization
or regularizations in coordinate space,
which are, to a large extent, equivalent to $q^2$-dependent
regulators in momentum space,
will lead to the same qualitative results.

For simplicity, we will use the same type of the regulator
for all parts of the LO and NLO potentials.
Moreover, the long-range and short-range parts of the LO potential
will be regularized using the same cutoff $\Lambda_0$. Similarly,  
the long-range and the leading short-range parts of the NLO potential
will be regularized by the same cutoff $\Lambda_2$. Other 
possible contact terms of the NLO potential are allowed
to have independent regulators $\Lambda_{2,i}$ with $i=2,4,\dots$.
The sharp cutoffs will be used in Sec.~\ref{Sec:LongYang}.

In general, the existence of an infinite-cutoff limit
implies that various cutoffs can be taken to infinity independently:
\begin{align}
 \Lambda_0(\tau)\xrightarrow[\tau\to\infty]{}\infty,\qquad
 \Lambda_2(\tau)\xrightarrow[\tau\to\infty]{}\infty,\qquad\dots,
 \label{Eq:trajectories}
\end{align}
where $\Lambda_0(\tau)$, $\Lambda_2(\tau)$,\dots are (in general) different functions.

The potential in the $^3P_0$ partial wave is obtained from the
potential in the plane wave basis
\begin{align}
V=& V_{C}+V_{\sigma} \vec{\sigma}_{1} \cdot \vec{\sigma}_{2}
+V_{\sigma q}\left(\vec{\sigma}_{1} \cdot \vec{q} \,
    \right)\left(\vec{\sigma}_{2} \cdot \vec{q} \, \right) 
+V_{S L} i \frac{1}{2}\left(\vec{\sigma}_{1}+\vec{\sigma}_{2}\right) \cdot(\vec{p} 
\times \vec{p}\, ')\,,
\label{Eq:structures}
\end{align}
where we kept only the structures relevant for our calculation, 
by the projection formula \cite{Epelbaum:1999dj}:
\begin{equation}
V(p',p)= 2 \pi \int_{-1}^{1} d z\Big\{z V_{C}(\vec p\,',\vec p \, )+z
V_{\sigma}(\vec p\,',\vec p \, )
-\left[\left(p^{\prime 2}+p^{2}\right) z-2 p^{\prime} p\right]
V_{\sigma q}(\vec p\,',\vec p \, )
+p' p (z^2-1)V_{SL}(\vec p\,',\vec p \, )\Big\}.
\label{Eq:PWprojection}
\end{equation}
The angular integration in Eq.~\eqref{Eq:PWprojection} is performed
over $z=\cos\theta$, with
$\theta$ the angle between the initial and final center-of-mass
momenta of the nucleons $\vec p$ and $\vec p\,'$.

The non-vanishing structures of the one-pion- and two-pion-exchange potentials
(up to irrelevant polynomial terms) read \cite{Kaiser:1997mw}:
\begin{align}
V_{1\pi,\sigma q}(\vec p\,',\vec p \, )= -\bigg(\frac{g_A}{2F_\pi}\bigg)^2  
\frac{1}{q^2 + M_\pi^2},
\label{Eq:V_1pi}
\end{align}
and
\begin{align}
V_{2\pi, C}(\vec p\,',\vec p \, )&= - \frac{L(q) }{384 \pi^2 F_\pi^4}  \biggl[4M_\pi^2 (5g_A^4 - 4g_A^2 -1)
+ q^2(23g_A^4 - 10g_A^2 -1)
+ \frac{48 g_A^4 M_\pi^4}{4 M_\pi^2 + q^2} \biggr],\nonumber\\
  V_{2\pi, \sigma q}(\vec p\,',\vec p \, )&=
                                            - \frac{1}{q^2}V_{2\pi,
                                            \sigma}(\vec p\,',\vec p
                                            \, ) =
-\frac{3 g_{A}^{4}}{64 \pi^{2} F_{\pi}^{4}}L(q),
\label{Eq:2pi_exchange}
\end{align}
where
\begin{align}
L(q)=\frac{1}{q} \sqrt{4 M_{\pi}^{2}+q^{2}} \log \frac{\sqrt{4 M_{\pi}^{2}+q^{2}}+q}{2 M_{\pi}}.
\end{align} 
Further, $M_\pi$, $F_\pi$ and $g_A$ refer to the pion mass, decay constant
and nucleon axial constant, respectively.
For illustrative purposes, we will also consider in Sec.~\ref{Sec:nonlocal} a simplified
version of the two-pion-exchange potential $\tilde V_{2\pi}(\vec
p\,',\vec p \,) $ by making it less singular
at short distances via
\begin{align}
\tilde V_{2\pi}(\vec p\,',\vec p\, )&=  V_{2\pi}(\vec p\,',\vec p \, )\frac{(3M_\pi)^2}{(3M_\pi)^2+q^2},
\label{Eq:2pi_exchange_tilde}
\end{align}
without modifying the longest-range part of the interaction.

The unregularized contact terms in the $^3P_0$ partial-wave have the
form 
\begin{align}
V^{\rm \scriptscriptstyle LO}_{\text{ct},0}&=V^{\rm \scriptscriptstyle NLO{}}_{\text{ct},0}=pp' ,\nonumber\\
V^{\rm \scriptscriptstyle NLO{}}_{\text{ct},2}&=pp'(p^2+p'^2) , \; \dots
\label{Eq:contact_terms}
\end{align}
When using local regulators, the
unregularized $^3P_0$ contact terms can be chosen via
\begin{align}
 V^{\rm \scriptscriptstyle LO}_{\text{ct},0,C}&=V^{\rm \scriptscriptstyle LO}_{\text{ct},0, \sigma}=-\frac{1}{32\pi}q^2,\nonumber\\
 V^{\rm \scriptscriptstyle LO}_{\text{ct},0,\sigma q}&=- \frac{1}{2} V^{\rm \scriptscriptstyle LO}_{\text{ct},0,SL}
                                                       =\frac{1}{16\pi},
 \label{Eq:contact_terms_local}
\end{align}
in order that projections onto other partial waves vanish, see also
Ref.~\cite{Gezerlis:2014zia}. 
Analogous expressions can be constructed for higher-order contact interactions.

The LO amplitude $T^{\rm \scriptscriptstyle LO}$ is obtained by solving the Lippmann-Schwinger equation
\begin{align}
T^{\rm \scriptscriptstyle LO}=V^{\rm \scriptscriptstyle LO}+V^{\rm \scriptscriptstyle LO}GT^{\rm \scriptscriptstyle LO},
\end{align}
or more explicitly,
\begin{align}
T^{\rm \scriptscriptstyle LO}(p',p;p_\text{on})&=V^{\rm \scriptscriptstyle LO}(p',p)
+\int \frac{p''^2 dp''}{(2\pi)^3}
V^{\rm \scriptscriptstyle LO}(p',p'')G(p'';p_\text{on})
T^{\rm \scriptscriptstyle LO}(p'',p;p_\text{on}),\nonumber\\
G(p''; p_\text{on})&=\frac{m_N}{p_\text{on}^2-p''^2+i \epsilon},
\label{Eq:LS_equation}
\end{align}
where $p_\text{on}$ denotes the on-shell center-of-mass momentum
and $m_N$ is the nucleon mass.
The NLO amplitude is given by the distorted-wave Born approximation
\begin{align}
T^{\rm \scriptscriptstyle NLO{}}=(\mathds{1}+T^{\rm \scriptscriptstyle LO}G) V^{\rm \scriptscriptstyle NLO{}} (\mathds{1}+G T^{\rm \scriptscriptstyle LO}).\label{Eq:T2}
\end{align}
We can also define separately $T_{2\pi}$ and $T_{\text{ct},i}$,
\begin{align}
 T_{2\pi}&=(\mathds{1}+T^{\rm \scriptscriptstyle LO}G)  V_{2\pi,\Lambda} (\mathds{1}+G T^{\rm \scriptscriptstyle LO}),\nonumber\\
 T_{\text{ct},i}&= (\mathds{1}+T^{\rm \scriptscriptstyle LO}G) V^{\rm \scriptscriptstyle NLO{}}_{\text{ct},i,\Lambda} (\mathds{1}+G T^{\rm \scriptscriptstyle LO}),\qquad i=0,2,\dots,
 \label{Eq:def_T_ct}
\end{align}
so that the full NLO amplitude is given by
\begin{align}
T^{\rm \scriptscriptstyle NLO{}}= T_{2\pi}+\sum_{i=0}^n C^{\rm \scriptscriptstyle NLO{}}_{2i} T_{\text{ct},2i}.
\end{align}
Notice further that the amplitude $T^{\rm \scriptscriptstyle LO}$ depends on the cutoff
$\Lambda_0$ alone,
while the amplitude $T^{\rm \scriptscriptstyle NLO{}}$ is a function of all cutoffs $\Lambda$,
which we do not always indicate explicitly.

\subsection{Dispersive approach}
As already mentioned in Sec.~\ref{Sec:Introduction}, 
the LO amplitude $T^{\rm \scriptscriptstyle LO}$ is required to have an infinite-cutoff
limit in the ``RG-invariant'' EFT scheme, see Eq.~\eqref{Eq:fininte_limit}.
Therefore, 
we assume that by choosing an appropriate renormalization condition 
for the constant $C_0=C_0(\Lambda_0)$,
one obtains a finite limit for the on-shell ($p'=p=p_\text{on}$)
LO amplitude:
\begin{align}
 T^{\rm \scriptscriptstyle LO}_{\Lambda}(p_\text{on})=T^{\rm \scriptscriptstyle LO}_{\infty}(p_\text{on})
 \{1+O[\left(\bar q/\Lambda_0\right)^{\alpha_{0}}]\},
\end{align}
where $\bar q=\max(p_\text{on},M_\pi)$ and $\alpha_{0}$ is some positive number.
This follows from the general theory of singular potentials \cite{Frank:1971xx}
and is supported by the numerical calculations performed in Ref.~\cite{Nogga:2005hy}.
Analogously, the phase shift $\delta^{\rm \scriptscriptstyle LO} (p_\text{on})$ approaches its 
limiting value at $\Lambda_0\to\infty$ for a given finite
value of $p_\text{on}$
when using the normalization 
$\delta (0)=0$
in order to avoid the shifts generated by deeply bound states.

As the renormalization condition, we are free to fix e.g.~the scattering volume
or the value of the phase shift at any energy point $p_0$, $|p_0|\ll\Lambda_0$, above or below threshold:
\begin{align}
 \delta^{\rm \scriptscriptstyle LO}(p_0)=\delta^{\rm \scriptscriptstyle LO}_0.
\end{align}
It is common to choose $\delta^{\rm \scriptscriptstyle LO}_0$ equal to the empirical 
phase shift extracted from experimental data:
\begin{align}
 \delta^{\rm \scriptscriptstyle LO}_0=\delta_\text{exp}(p_0).
 \label{Eq:condition_delta_0_exp}
\end{align}

In the considerations below we assume 
that we can regard both $V^{\rm \scriptscriptstyle LO}$ and $V^{\rm \scriptscriptstyle NLO{}}$ as being local.
Strictly speaking, this is only the case if the regulators of the LO and NLO potentials are local.
If the regulators are non-local, one can still expect our arguments to
apply, because the non-localities appear at momenta $p\sim\Lambda$ and
should only induce
corrections vanishing in the $\Lambda\to\infty$ limit, which are
allowed in the resulting formulas in any case.
Since the potentials are assumed to be local, one can use an $N/D$ representation for the amplitude $T^{\rm \scriptscriptstyle LO}$~\cite{Newton:1982qc}
\begin{align}
 T^{\rm \scriptscriptstyle LO}(p_\text{on})=\frac{p_\text{on}^2 \tilde N^{\rm \scriptscriptstyle LO}(p_\text{on})}{\tilde D(p_\text{on})},
 \label{Eq:NoverD_tilde}
\end{align}
where $\tilde D(p_\text{on})$ is the Fredholm determinant (equal to the Jost function)
and contains all right-hand singularities and bound-state poles of $T^{\rm \scriptscriptstyle LO}$.
It can be represented by means of the dispersion relation~\cite{Newton:1982qc}
\begin{align}
 \tilde D(p_\text{on})=
 \Bigg( \prod_{j=1}^n \frac{p_\text{on}^2+\kappa_j^2}{p_\text{on}^2} \Bigg)
 \exp\bigg[\frac{1}{\pi}
 \int_0^{\infty}  dp^2 \frac{\tilde\delta^{\rm \scriptscriptstyle LO}(p)}{p_\text{on}^2-p^2+i\epsilon}
 \bigg],
 \label{Eq:DR_D_tilde}
\end{align}
where $n$ denotes the number of bound states and $p_j=i\kappa_j$ are the positions of the bound-state poles.
The factor $p_\text{on}^2$ in front of $\tilde N^{\rm \scriptscriptstyle LO}$ in Eq.~\eqref{Eq:NoverD_tilde}
reflects the orbital angular momentum $l=1$.
In Eq.~\eqref{Eq:DR_D_tilde}, $\tilde \delta^{\rm \scriptscriptstyle LO}(p)$ is the LO phase shift normalized
to $\tilde \delta^{\rm \scriptscriptstyle LO}(\infty)=0$.
In our case, it is more suitable to use an alternative representation with 
a subtraction at $p_\text{on}=0$ in terms of $\delta^{\rm \scriptscriptstyle LO}(p)$:
\begin{align}
 D(p_\text{on})=
 \Bigg( \prod_{j=1}^n \frac{p_\text{on}^2+\kappa_j^2}{\kappa_j^2} \Bigg)
 \exp\left[\frac{1}{\pi}
 \int_0^{\infty}  dp^2\frac{p_\text{on}^2}{p^2} \frac{\delta^{\rm \scriptscriptstyle LO}(p)}{p_\text{on}^2-p^2+i\epsilon}
 \right],
 \label{Eq:DR_D}
\end{align}
so that
\begin{align}
 T^{\rm \scriptscriptstyle LO}(p_\text{on})=\frac{p_\text{on}^2 N^{\rm \scriptscriptstyle LO}(p_\text{on})}{D(p_\text{on})},
\end{align}
where $N^{\rm \scriptscriptstyle LO}(p_\text{on})$ contains only left-hand singularities.
Under the reasonable assumption that the integral in Eq.~\eqref{Eq:DR_D}
converges at momenta $p\ll\Lambda_0$, 
and provided the bound states lie far away from the
threshold\footnote{In fact, it was indicated in
  Ref.~\cite{Nogga:2005hy} that the positions of the bound states
approach fixed values for $\Lambda_0\to \infty$.},
we can conclude that $D(p_\text{on})$
also approaches a finite limit as $\Lambda_0\to\infty$:
\begin{align}
 D_{\Lambda_0}(p_\text{on})=D_\infty(p_\text{on})\{1+O[\left(\bar q/\Lambda_0\right)^{\alpha_D}]\},
 \qquad \alpha_D>0.
 \label{Eq:D_infty}
\end{align}

The $N/D$ representation for the subleading amplitudes $T_{2\pi}$ and $T_{\text{ct},i}$ reads
\begin{align}
T_{2\pi}(p_\text{on})&=\frac{p_\text{on}^2 N_{2\pi}(p_\text{on})}{D(p_\text{on})^2},\nonumber\\
T_{\text{ct},i}(p_\text{on})&=\frac{p_\text{on}^2 N_{\text{ct},i}(p_\text{on})}{D(p_\text{on})^2}.
\end{align}
Again, the functions $N_{2\pi}(p_\text{on})$ and $N_{\text{ct},i}(p_\text{on})$ possess only left-hand singularities.
If we further assume that both the LO and the NLO potentials are of
Yukawa type,
which is true for the regulators of the form given in Eq.~\eqref{Eq:local_regulator2},
the location of the left-hand singularities 
and their properties are well known \cite{Martin:1961}, and we can write down the 
corresponding dispersion relations
\begin{align}
 N_{2\pi}(p_\text{on})&=
 \int_{-\infty}^{-M_\pi^2} \frac{dp^2}{\pi} \frac{\Delta N_{2\pi}(p)}{p^2-p_\text{on}^2},\nonumber\\
 N_{\text{ct},i}(p_\text{on})&=
 \int_{-\infty}^{-\Lambda_{2,i}^2/4} \frac{dp^2}{\pi} \frac{\Delta N_{\text{ct},i}(p)}{p^2-p_\text{on}^2},
 \qquad \Lambda_{2,0}\equiv\Lambda_2.
 \label{Eq:DR_N2pi_ct}
\end{align}
The discontinuity $\Delta N_{2\pi}(p)$ in the region $-\Lambda_\text{min}^2/4 < p^2<-M_\pi^2$,
where $\Lambda_\text{min}=\min(\Lambda)$,
is determined by the perturbative contributions such as $V^{\rm \scriptscriptstyle NLO{}}$, $V^{\rm \scriptscriptstyle NLO{}} G V^{\rm \scriptscriptstyle LO}$, etc., and
is independent of $\Lambda$.
Therefore, we can expect the following $\Lambda$-behaviour of $N_{2\pi}$ and $N_{\text{ct},i}$:
\begin{align}
 N_{2\pi,{\Lambda}}(p_\text{on})&=
 N_{2\pi,\infty}(p_\text{on})+P_{2\pi,{\Lambda}}(p_\text{on}^2)+O[\left(\bar q/\Lambda\right)^{\alpha_{2\pi}}],\nonumber\\
 N_{\text{ct},i,{\Lambda}}(p_\text{on})&=P_{\text{ct},i,{\Lambda}}(p_\text{on}^2)+O[\left(\bar q/\Lambda\right)^{\alpha_{\text{ct}}}],
 \qquad \alpha_{2\pi},\alpha_{\text{ct}}>0,
 \label{E:N_infty}
\end{align}
where $P_{2\pi,{\Lambda}}(p_\text{on}^2)$ and $P_{\text{ct},i,{\Lambda}}(p_\text{on}^2)$
are polynomials in $p_\text{on}^2$, which may contain positive powers of $\Lambda$,
and $\bar q/\Lambda$ denotes $\bar q/\Lambda_\text{min}$.
Their degrees correspond to the number of subtractions necessary to suppress the 
momentum region $p\sim\Lambda$ in dispersive integrals in Eq.~\eqref{Eq:DR_N2pi_ct}.

Now, let us first assume that a single subtraction is sufficient, i.e.,
both  polynomials $P_{2\pi,{\Lambda}}(p_\text{on}^2)$ and
$P_{\text{ct},i,{\Lambda}}(p_\text{on}^2)$  are simply $\Lambda$-dependent constants,
\begin{equation}
 P_{2\pi,{\Lambda}}(p_\text{on}^2) \equiv C_{2\pi}({\Lambda}),\quad
 \quad 
 P_{\text{ct},0,{\Lambda}}(p_\text{on}^2) \equiv C_{\text{ct}}({\Lambda}),
\end{equation}
and
\begin{align}
 C^{\rm \scriptscriptstyle NLO{}}_{i}=0,\text{ for }i\ne 0.
\end{align}
This situation describes the case when
a more regular
version of the two-pion-exchange potential defined in
Eq.~\eqref{Eq:2pi_exchange_tilde} is used.
We will study such a simplified model numerically in the
next subsection.

Combining the estimates in Eqs.~\eqref{Eq:D_infty} and~\eqref{E:N_infty}
we obtain the following expression for the NLO amplitude:
\begin{align}
 T^{\rm \scriptscriptstyle NLO{}}_{{\Lambda}}(p_\text{on})&=\frac{p_\text{on}^2}{D_\infty(p_\text{on})^2}
  \Big\{N_{2\pi,\infty}(p_\text{on})+C_{2\pi}({\Lambda})+C^{\rm \scriptscriptstyle NLO{}}_{0}({\Lambda})
   \left[C_{\text{ct}}({\Lambda})+\delta N_{\text{ct},{\Lambda}}(p_\text{on})\right]
  +\delta N_{2\pi,{\Lambda}}(p_\text{on})\Big\},\nonumber\\
   \delta N_{2\pi,{\Lambda}}(p_\text{on})&=O[\left(\bar q/\Lambda\right)^{\alpha}],\nonumber\\
   \delta N_{\text{ct},{\Lambda}}(p_\text{on})&=O[\left(\bar q/\Lambda\right)^{\beta}],
   \label{Eq:T2_Lambda}
\end{align}
with $\alpha,\beta>0$.
We can now fix the constant $C^{\rm \scriptscriptstyle NLO{}}_{0}({\Lambda})$ by choosing some renormalization
condition. If
the LO amplitude is fixed by the requirement to reproduce the experimental value
at $p_\text{on}=p_0$, the natural choice is to set
\begin{align}
 T^{\rm \scriptscriptstyle NLO{}}(p_0)=0.
 \label{Eq:condition_T2}
\end{align}
Then, if one can neglect $\delta N_{\text{ct},{\Lambda}}$ and $\delta N_{2\pi,{\Lambda}}$,
the solution to Eq.~\eqref{Eq:condition_T2} is
\begin{align}
 C^{\rm \scriptscriptstyle NLO{}}_{0}({\Lambda})=-\frac{N_{2\pi,\infty}(p_0)+C_{2\pi}({\Lambda})}{C_{\text{ct}}({\Lambda})}, 
 \label{Eq:C_2_Lambda}
\end{align}
and the NLO amplitude indeed becomes $\Lambda$-independent as $\Lambda$ tends to infinity:
\begin{align}
 T^{\rm \scriptscriptstyle NLO{}}_{{\Lambda}}(p_\text{on})&\approx\frac{p_\text{on}^2}{D_\infty(p_\text{on})^2}
  \Big[N_{2\pi,\infty}(p_\text{on})-N_{2\pi,\infty}(p_0)\Big].
\end{align}

However, the term involving $\delta N_{\text{ct},{\Lambda}}$ can \emph{not} always be neglected
since it is multiplied by $C^{\rm \scriptscriptstyle NLO{}}_{0}({\Lambda})$ which, in general, is
unbounded as follows from Eq.~\eqref{Eq:C_2_Lambda}.
As follows from the definition of $T_{\text{ct}, 0}$ in Eq.~\eqref{Eq:def_T_ct},
the constant $C_{\text{ct}}({\Lambda})$
can be viewed as being
proportional to the square of the LO scattering wave function, smeared with some weight
over the short distance region $r\sim 1/\Lambda_2$, where it strongly oscillates
as $\Lambda_0$ increases\footnote{This is a characteristic feature of
  solutions to singular potentials \cite{Frank:1971xx}.}.
Therefore, for some values of $\Lambda_0$ and $\Lambda_2$, 
$C_{\text{ct}}({\Lambda})$ can become very small, and Eq.~\eqref{Eq:C_2_Lambda}
must be replaced with
\begin{align}
 C^{\rm \scriptscriptstyle NLO{}}_{0}({\Lambda})=-\frac{N_{2\pi,\infty}(p_0)+C_{2\pi}({\Lambda})}{C_{\text{ct}}({\Lambda})+\delta N_{\text{ct},{\Lambda}}(p_0)}. 
 \label{Eq:C_2_Lambda_2}
\end{align}
As one can see, if for some ``exceptional'' value $\Lambda=\bar\Lambda$, one has
\begin{align}
 C_{\text{ct}}({\bar\Lambda})+\delta N_{\text{ct},{\bar\Lambda}}(p_0)=0,
 \label{Eq:condition1}
\end{align}
while
\begin{align}
  C_{\text{ct}}({\bar\Lambda})+\delta N_{\text{ct},{\bar\Lambda}}(p_\text{on})\not\equiv 0,
  \label{Eq:condition2}
\end{align}
i.e., if the zero is not factorizable,
then the cutoff independence of $T^{\rm \scriptscriptstyle NLO{}}$ in the vicinity of 
$\bar\Lambda$ and therefore
also in general may become questionable.
If the constant $C^{\rm \scriptscriptstyle NLO{}}_{0}({\Lambda})$ behaves near the pole $\Lambda=\bar\Lambda$ as
\begin{align}
 C^{\rm \scriptscriptstyle NLO{}}_{0}({\Lambda})\approx \bar C^{\rm \scriptscriptstyle NLO{}}_{0}(\bar\Lambda)(\Lambda-\bar\Lambda)^{-\bar\alpha},
\end{align}
where typically $\bar\alpha = 1$ or $\bar\alpha = 2$, see the next sections,
then, as follows from Eq.~\eqref{Eq:T2_Lambda},
the width $\delta \Lambda$ of the ``exceptional'' regions
is roughly of order
\begin{align}
\delta \Lambda \sim 
\frac{\bar C^{\rm \scriptscriptstyle NLO{}}_{0}(\bar\Lambda)^{1/\bar\alpha}}{\bar\Lambda^{\beta/\bar\alpha}},
\label{Eq:deltaLambda}
\end{align}
and generally decreases with $\Lambda$ remaining, nevertheless, finite.
In the next two subsections we will see how this situation actually occurs in
the case of non-locally and locally regulated potentials.

The above discussion was merely meant to identify the 
``exceptional'' values of the cutoffs that may destroy 
the cutoff independence of the amplitude. In the case of a single subtraction, 
the relation in Eq.~\eqref{Eq:C_2_Lambda_2} is equivalent to
\begin{equation}
  T^{\rm \scriptscriptstyle
    NLO{}}_{\text{ct},0,\bar\Lambda}(p_0)=0,\quad \quad
  T^{\rm \scriptscriptstyle NLO{}}_{\text{ct},0,\bar\Lambda}(p_\text{on})\not\equiv 0.
\label{Eq:Lambda_bar_T2_0}
  \end{equation}
If more than one, say $n_\text{sub}$, subtractions (NLO counter terms) are necessary,
we will need additional renormalization conditions to fix the corresponding LECs.
It is convenient to 
define a function $\delta^{\rm \scriptscriptstyle NLO{}}(p_\text{on})$,
\begin{align}
  T^{\rm \scriptscriptstyle NLO{}}(p_\text{on})=-\delta^{\rm
  \scriptscriptstyle NLO{}}(p_\text{on}) \, e^{2i\delta^{\rm \scriptscriptstyle LO}(p_\text{on})}\rho(p_\text{on}),
  \label{Eq:delta2_definition}
\end{align}
with the phase space factor 
\begin{align}
\rho(p_\text{on})=\frac{m_N}{(2\pi)^3}\frac{\pi p_\text{on}}{2}. 
\end{align}
Analogously, we  define the functions $\delta_{2\pi}(p_\text{on})$ and
 $\delta_{\text{ct},i}(p_\text{on})$:
\begin{align}
  T_{2\pi}(p_\text{on})&=-\delta_{2\pi}(p_\text{on}) \, e^{2i\delta^{\rm \scriptscriptstyle LO}(p_\text{on})}\rho(p_\text{on}),\nonumber\\
  T_{\text{ct},i}(p_\text{on})&=-\delta_{\text{ct},i}(p_\text{on}) \, e^{2i\delta^{\rm \scriptscriptstyle LO}(p_\text{on})}\rho(p_\text{on}),
  \qquad i=0,2,\dots,2(n_\text{sub}-1).
  \label{Eq:delta2pi_ct_definition}
\end{align}
The functions $\delta^{\rm \scriptscriptstyle NLO{}}(p_\text{on})$, $\delta_{2\pi}(p_\text{on})$ and
$\delta_{\text{ct},i}(p_\text{on})$  are real-valued in the elastic
physical region due to the unitarity of the $S$-matrix.
They can be identified with the perturbative NLO correction to the phase shift
as done e.g.~in Ref.~\cite{Long:2011qx}.
One can impose the following renormalization conditions:
\begin{align}
 \delta^{\rm \scriptscriptstyle NLO{}}(p_i)=\delta^{\rm
  \scriptscriptstyle NLO{}}_i,\qquad i=0,\, 2,\, \dots, \, 2(n_\text{sub}-1),
\end{align}
where $p_0$ can be chosen to be the same on-shell momentum as used
in the renormalization of the LO amplitude.
Alternatively, one can use derivatives of $\delta^{\rm \scriptscriptstyle NLO{}}(p_\text{on})$ with respect
to $p_\text{on}^2$ at threshold or impose other conditions.
In turn, one can use the empirical phase shifts and set
\begin{align}
  \delta^{\rm \scriptscriptstyle NLO{}}_i=\delta_\text{exp}(p_i)- \delta^{\rm \scriptscriptstyle LO}(p_i),
  \label{Eq:condition_delta_2_exp}
\end{align}
which yields $\delta^{\rm \scriptscriptstyle NLO{}}_0=0$ if the condition~\eqref{Eq:condition_delta_0_exp} is used.
By analogy with Eq.~\eqref{Eq:Lambda_bar_T2_0}, we can easily identify the ``exceptional'' cutoff values $\bar\Lambda$
lying on the trajectories $\Lambda(\tau)$ defined in Eq.~\eqref{Eq:trajectories}
that may destroy the cutoff independence of the NLO amplitude via
\begin{equation}
 \det A_{\bar\Lambda}=0,\quad \quad
 \det \tilde A_{\bar\Lambda}\ne0,
 \label{Eq:condition_determinant}
\end{equation}
with ($\Lambda$ dependence being omitted)
\begin{equation}
 A_{ij}=\delta_{\text{ct},i}(p_j),\qquad
 \tilde A_{ij}=\delta_{\text{ct},i}(\tilde p_j),\qquad i,j=0,\, 2,\,
 \dots,\, 2(n_\text{sub}-1),
\end{equation}
where at least for one $j$, $\tilde p_j\ne p_j$.
In what follows, we will present
numerical evidence
that such ``exceptional'' cutoff values cannot be avoided in general.
The case of a single subtraction will be considered in Sec.~\ref{Sec:nonlocal},
while the case of two subtractions will be discussed in Sec.~\ref{Sec:LongYang}.

In all calculations presented below, the following numerical values for the physical constants are employed:
the pion decay constant is set to $F_\pi=92.1$ MeV,
the isospin average nucleon and pion masses
are $m_N=938.9$~MeV and $M_\pi=138.04$~MeV, respectively, 
and the effective axial coupling constant of the nucleon 
is set to $g_A = 1.29$ to account for 
the Goldberger-Treiman discrepancy, see e.g.~Refs.~\cite{Epelbaum:2004fk, Fettes:1998ud}.
The calculations have been performed using \emph{Mathematica} \cite{mathematica12.0}.

\subsection{Simplified model with a non-local regulator}
\label{Sec:nonlocal}
In this subsection we consider the simplified model for the $^3P_0$
channel using a modified $2\pi$-exchange introduced in Sec.~\ref{Sec:formalism}
with the non-local regulator at LO and NLO, see Eq.~\eqref{Eq:nonlocal_regulator}.
In this scheme only one subtraction at NLO is needed. We fix the constants $C^{\rm \scriptscriptstyle LO}_0$ and $C^{\rm \scriptscriptstyle NLO{}}_0$
by the renormalization conditions
\begin{equation}
 \delta^{\rm \scriptscriptstyle LO}(p_0)=\delta_\text{exp}(p_0),\qquad
  \delta^{\rm \scriptscriptstyle NLO{}}(p_0)=0,
  \label{Eq:conditions_p_0}
\end{equation}
where the on-shell momentum $p_0$ is chosen to correspond 
to the laboratory energy of $T_\text{lab}=50$~MeV.

The contact part of the on-shell NLO amplitude
(Eq.~\eqref{Eq:def_T_ct}) takes the particularly simple form
 \begin{align}
 T_{\text{ct}}(p_\text{on})&=C^{\rm \scriptscriptstyle NLO{}}_0 T_{\text{ct},0}(p_\text{on})=C^{\rm \scriptscriptstyle NLO{}}_0 \psi_\Lambda(p_\text{on})^2,
 \label{Eq:T_ct_nonlocal}
\end{align}
with the vertex function
\begin{align}
 &\psi_\Lambda(p_\text{on}) =p_\text{on}F_{\Lambda_2}(p_\text{on})
   +\int \frac{p^2 dp}{(2\pi)^3}p\, F_{\Lambda_2}(p)\,
   G(p;p_\text{on})\, 
 T^{\rm \scriptscriptstyle LO}_{\Lambda_0}(p, p_\text{on};p_\text{on}).
 \end{align}
Therefore, the $\Lambda$-independence of the NLO amplitude can be 
potentially destroyed if 
\begin{align}
\psi_\Lambda(p_0)=0
\label{Eq:psi_Lambda_equals_0}
\end{align}
for some values of $\Lambda_0$ and $\Lambda_2$.

We start with considering a special case with the LO and NLO cutoffs
being set to the same values, $\Lambda_2=\Lambda_0$,
and show that under this condition, the zero of $\psi_\Lambda(p_\text{on})$
is factorizable.
One can demonstrate this using the two-potential formalism.
The off-shell LO amplitude is represented as
 \begin{align}
  T^{\rm \scriptscriptstyle LO}_{\Lambda_0}(p',p;p_\text{on})=T_{1\pi,\Lambda_0}(p',p;p_\text{on})
  +\frac{\psi_{1\pi,\Lambda_0}(p';p_\text{on}) \, \psi_{1\pi,\Lambda_0}(p;p_\text{on})}{C_0^{-1}-\Sigma_{\Lambda_0}(p_\text{on})},
 \end{align}
 where 
  \begin{align}
 \psi_{1\pi,\Lambda_0}(p;p_\text{on})& =p F_{\Lambda_0}(p)+\int \frac{p'^2 dp'}{(2\pi)^3}p'\, F_{\Lambda_0}(p')\, G(p';p_\text{on})
 \, T_{1\pi,\Lambda_0}(p', p;p_\text{on}),\nonumber\\
\Sigma_{\Lambda_0}(p_\text{on})&=\int \frac{p^2 dp}{(2\pi)^3}p\, F_{\Lambda_0}(p)G(p;p_\text{on})\psi_{1\pi,\Lambda_0}(p;p_\text{on}).
 \end{align}
 The $T$-matrix $T_{1\pi}$ is the solution of the LO Lippmann-Schwinger equation 
 without the contact term:
  \begin{align}
  T_{1\pi}=V_{1\pi}+V_{1\pi}G T_{1\pi}.
 \end{align}
It is straightforward to see that
\begin{align}
 \psi_{\Lambda_0}(p_\text{on})=\frac{\psi_{1\pi,\Lambda_0}(p_\text{on})}{1-C_0\Sigma_{\Lambda_0}(p_\text{on})},
\end{align}
where $\psi_{1\pi,\Lambda_0}(p_\text{on})\coloneqq \psi_{1\pi,\Lambda_0}(p_\text{on};p_\text{on})$.

Numerical checks show that the condition $\psi_{1\pi,\Lambda_0}(p_\text{on})=0$
is never fulfilled in the physical region.
Therefore, $\psi_{\Lambda_0}(p_\text{on})$ vanishes only when
$C_0=\infty$. 
This happens for the ``exceptional'' values of $\Lambda_0=\bar\Lambda_i$, $i=1,2,\dots$, due to the limit-cycle-like behaviour of $C_0(\Lambda_0)$,
see, e.g.~Ref.~\cite{Nogga:2005hy}.
In this case, $\psi_{\Lambda_0}(p_\text{on})$ vanishes identically for all energies as
 \begin{align}
 \psi_{\Lambda_0}(p_\text{on})\sim (\Lambda_0-\bar\Lambda_i)\xi_{\bar\Lambda_i}(p_\text{on}).
 \label{Eq:zeros_factorization}
\end{align}
Then, the constant $C^{\rm \scriptscriptstyle NLO{}}_0$ in the vicinity of $\bar\Lambda_i$
behaves as 
 \begin{align}
 C^{\rm \scriptscriptstyle NLO{}}_0\sim (\Lambda_0-\bar\Lambda_i)^{-2},
\end{align}
canceling the overall prefactor $(\Lambda_0-\bar\Lambda_i)^2$ in $T_{\text{ct}}$,
and no problems due to an enhancement of $1/\Lambda^\beta$-terms (see the discussion in the previous subsection
and Eq.~\eqref{Eq:T2_Lambda}) occur.

\begin{figure}[tb]
\includegraphics[width=\textwidth]{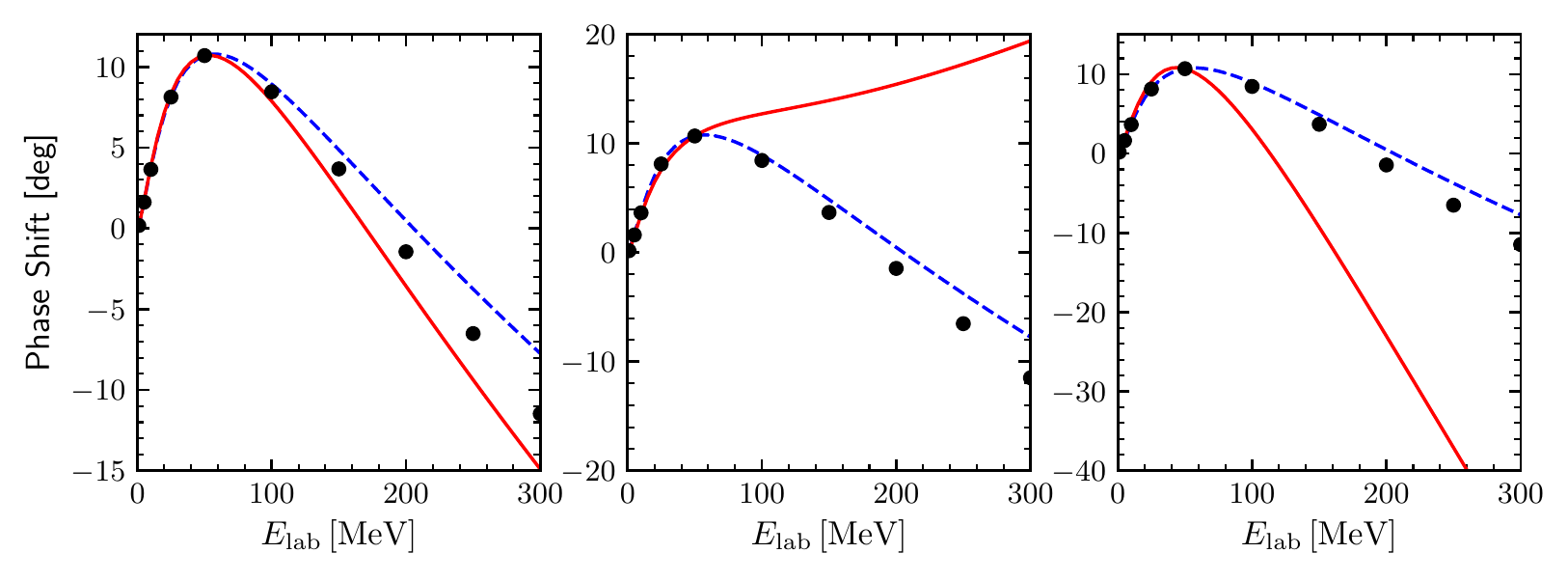}
\caption{Nucleon-nucleon phase shifts for the $^3P_0$ partial wave
  versus the laboratory energy
  calculated using 
the simplified model with a modified two-pion-exchange potential from Eq.~(\ref{Eq:2pi_exchange_tilde}).
Dashed  and solid lines denote the LO and NLO results, respectively.
Solid dots are the empirical phase shifts from the Nijmegen partial wave analysis~\cite{Stoks:1993tb}.
The left panel corresponds to a typical value of the cutoff, whereas the other two panels
show the phase shifts for the cutoff values near the ``exceptional''
ones as described in the text.
Plots are created using Matplotlib \cite{Hunter:2007}.} \label{Fig:ToyModel_phase_shifts}
\end{figure}

The phase shifts corresponding to this solution for a sufficiently large cutoff are shown
in the left panel of Fig.~\ref{Fig:ToyModel_phase_shifts}.
\begin{figure}[tb]
\includegraphics[width=0.95\textwidth]{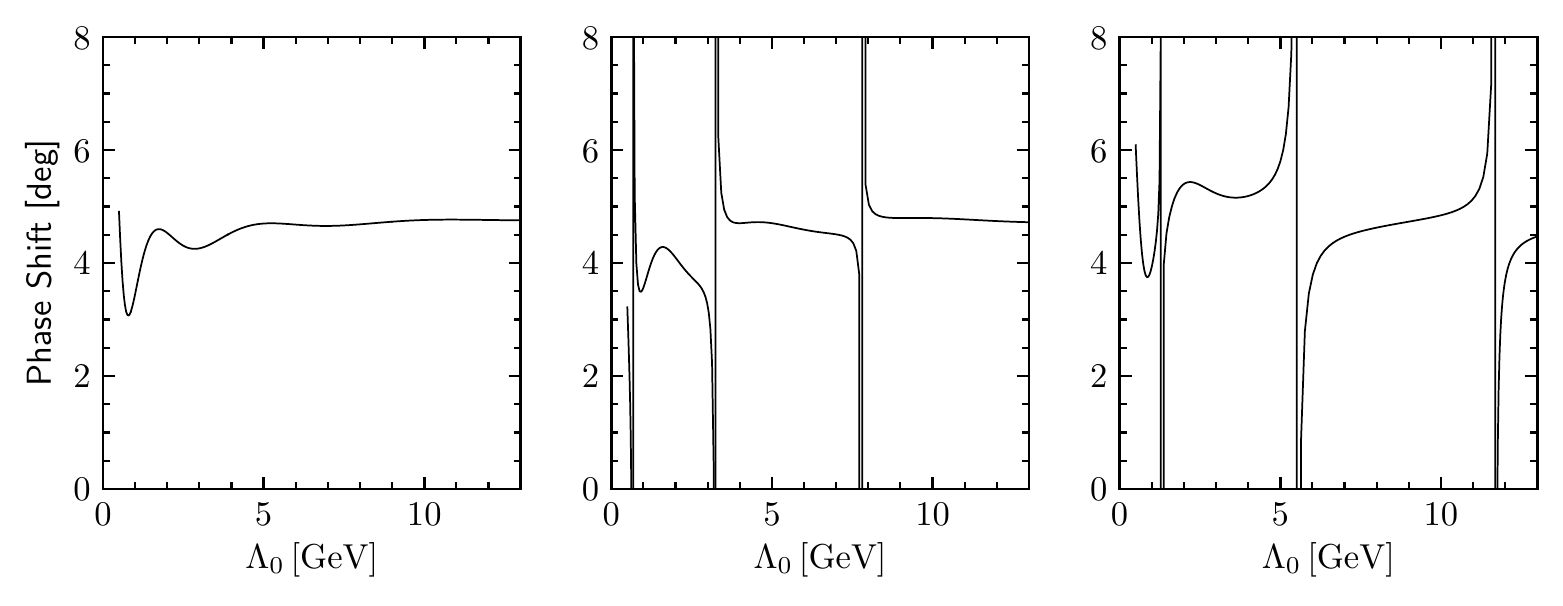}
\caption{The $^3P_0$ phase shift at the fixed laboratory energy of $T_\text{lab}=130$~MeV 
calculated in the simplified model at NLO as a function of the cutoff for
$\Lambda_2=\Lambda_0$ (left panel), $\Lambda_2=2\Lambda_0$ (middle panel)
and $\Lambda_2=\Lambda_0/2$ (right
panel).} \label{Fig:ToyModel_phase_shift_130}
\end{figure}
In the left panel of Fig.~\ref{Fig:ToyModel_phase_shift_130},
the NLO result for the phase shift $\delta^{\rm \scriptscriptstyle LO}+\delta^{\rm \scriptscriptstyle NLO{}}$, where $\delta^{\rm \scriptscriptstyle NLO{}}$ is defined 
according to Eq.~\eqref{Eq:delta2_definition}, is shown as a function of the
cutoff at the laboratory energy $T_\text{lab}=130$~MeV.
One indeed observes the cutoff-independent result as $\Lambda \to \infty$.
The situation changes, however,  if one sets $\Lambda_2\ne\Lambda_0$. 
Below, we consider two cases of a linear dependence of $\Lambda_2$ on $\Lambda_0$,
$\Lambda_2=2\Lambda_0$ and $\Lambda_2=\Lambda_0/2$, which reflect a typical situation
for any other kind of such a dependence.

\begin{figure}[tb]
\includegraphics[width=\textwidth/2]{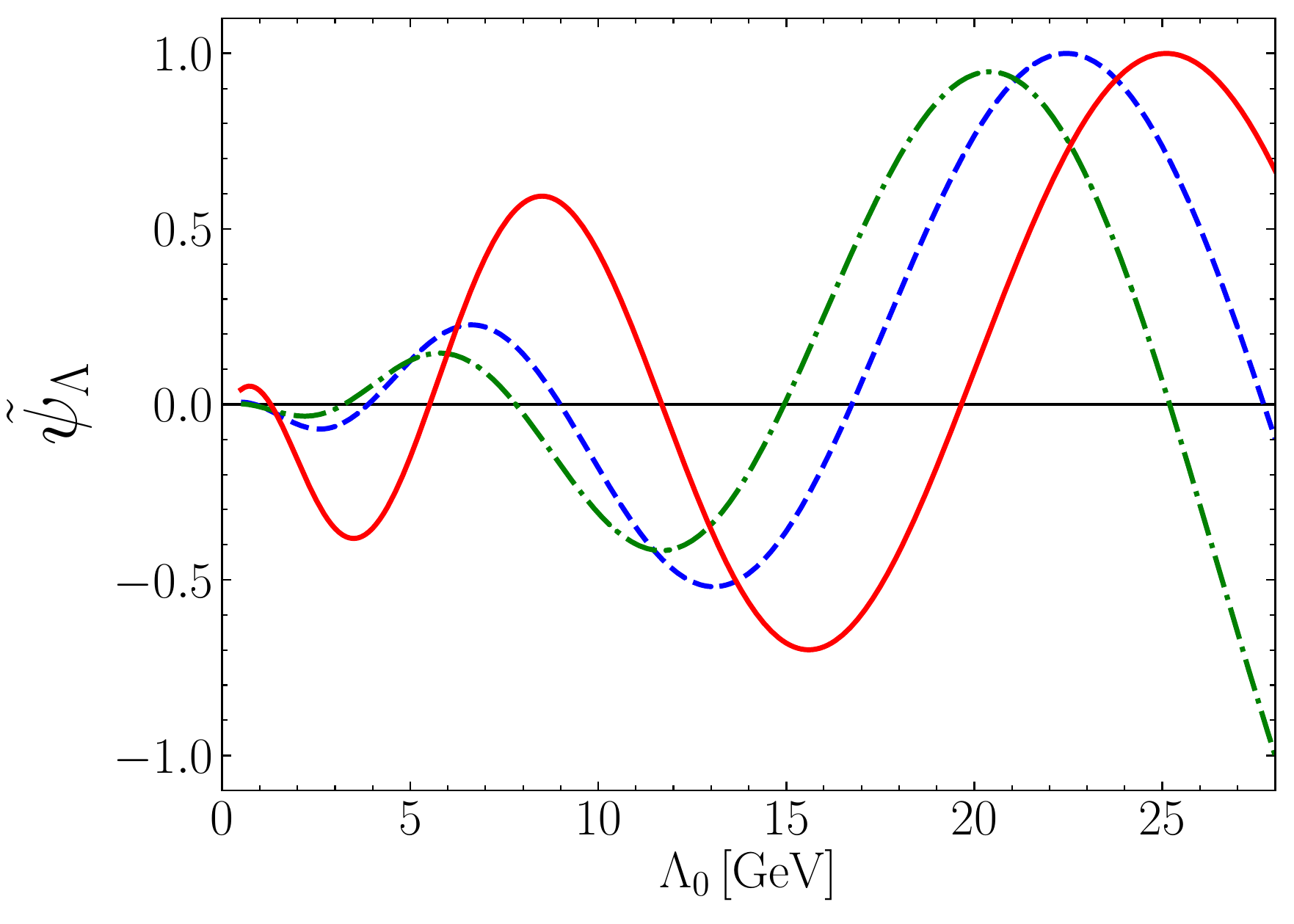}
\caption{The magnitude of the vertex function $\tilde\psi_{\Lambda}(p_\text{on})$
defined in Eq.~(\ref{DefTemp1}) for the simplified model at $p_\text{on}=p_0$.
Dashed, dash-dotted and solid lines correspond to 
the choice $\Lambda_2=\Lambda_0$, $\Lambda_2=2\Lambda_0$ and $\Lambda_2=\Lambda_0/2$, respectively.} \label{Fig:ToyModel_VertexFunction}
\end{figure}

Figure~\ref{Fig:ToyModel_VertexFunction} shows the locations of zeros of the 
magnitude of the vertex function $\tilde\psi_{\Lambda}(p_\text{on})$, defined as
\begin{align}
  \label{DefTemp1}
 \psi_{\Lambda}(p_\text{on})=\tilde\psi_{\Lambda}(p_\text{on})\, e^{i\delta^{\rm \scriptscriptstyle LO}}.
\end{align}
These zeros extend to $\Lambda=\infty$ and for $\Lambda_2\ne\Lambda_0$ correspond to ``exceptional''
values of the cutoff\footnote{The exact location of the zeros depends on the details of the LO potential.}.
The middle and right panels of Fig.~\ref{Fig:ToyModel_phase_shift_130}
demonstrate that the NLO phase shifts reveal a behaviour far from being constant 
in the vicinity of the ``exceptional'' cutoffs.
The cutoff dependence of the phase shift is more complicated than
just a double pole. In the case of $\Lambda_2=2\Lambda_0$ ($\Lambda_2=\Lambda_0/2$)
slightly above (below) the pole, there is a point with
$\psi_{\Lambda}(p_\text{on})=0$ (where $p_\text{on}\ne p_0$ corresponds to the considered $T_\text{lab}=130$~MeV).
At this point the NLO amplitude is given completely by the two-pion-exchange term,
which leads to another (finite) oscillation of the phase shift in the positive direction.
This is a characteristic feature of the scheme with one subtraction.

In Fig.~\ref{Fig:ToyModel_phase_shifts},
we show how the phase shifts deviate from the solution for the 
typical cutoff, plotted in the left panel, 
for the ``exceptional'' cutoff value $\bar\Lambda_0=19660.9$~MeV
using $\Lambda_2=\Lambda_0/2$:
the middle (right) panel corresponds to the choice $\Lambda_0=\bar\Lambda_0-5$~MeV ($\Lambda_0=\bar\Lambda_0+5$~MeV).

Note that the first ``exceptional'' cutoff values appear at $\Lambda_0\sim700-1300$~MeV,
i.e.~for $\Lambda_0$ of the order of or slightly above the hard scale $\Lambda_b$.  This does not
automatically imply that
a reasonable description of the data is impossible for such cutoff values.
For $\Lambda_0\sim\Lambda_b$,
the numerical value of the amplitude $T_{2\pi}$ is rather small
in line with expectations based on naive dimensional analysis.
Therefore, one could replace the renormalization condition  \eqref{Eq:conditions_p_0}, which is the 
main source of the problem, by the condition 
\begin{align}
 \delta^{\rm \scriptscriptstyle LO}(p_0)+\delta^{\rm \scriptscriptstyle NLO}(p_0)&=\delta_\text{exp}(p_0),\nonumber\\
 C^{\rm \scriptscriptstyle NLO{}}_{0}&=0.
 \label{Eq:condition_C20_0}
\end{align}
However, in doing so, one unavoidably violates one of the basic principles
of the infinite-cutoff scheme stating that
the cutoff independence must be achieved at each order individually.
Obviously, the condition~\eqref{Eq:condition_C20_0} cannot be 
adopted for $\Lambda\gg\Lambda_b$, because
in that case
the two-pion-exchange amplitude
would violate
the dimensional power counting if no subtractions are performed.
We do not claim though that no alternative  renormalization
schemes can be defined, which would employ renormalization conditions
different from those in Eq.~\eqref{Eq:conditions_p_0}, yet
approaching them in the limit $\Lambda\to\infty$ with no ``exceptional''
cutoffs. However, in such a scheme, the renormalization condition for the LO
amplitude would depend on a choice of the regulator of both the LO and the NLO potentials,
which seems to make no sense from the physical point of view.

A comment is in order here regarding the choice $\Lambda_2=\Lambda_0$ discussed above,
which prevents the appearance of ``exceptional'' cutoff values for the
considered interactions.  One might try to argue that
this is a ``natural'' choice, since the LO contact term and the 
$p' p$ NLO contact terms are parts of the same term in the effective Lagrangian
that is split between two orders.
However, insisting on such a choice contradicts the principle of the infinite-cutoff scheme
that the limit $\Lambda\to\infty$  should not depend on the functional
form of the regulator.
Note that in the limit $\Lambda\to\infty$, the two contact interactions indeed become identical
independently of the relationship between $\Lambda_0$ and $\Lambda_2$.
Moreover, this argument loses its justification completely
when additional contact interactions at NLO are included, as 
there is no physical reason to relate the regulators of such contact terms with the leading-order one.
One can in this case not even unambiguously define, what
the notion of the same regulator actually means
due to a different short-range structure of the contact interactions.
In particular, choosing the same sharp cutoff in momentum space 
still leads to the appearance of  ``exceptional'' values, as we will see in Sec.~\ref{Sec:LongYang}.
The only possibility to avoid the ``exceptional'' values as far as we
can see is to allow for energy-dependent contact terms of the form
$pp' p_\text{on}^{2i}$, $i=0,1,\dots$ only and to 
employ a unified form factor in momentum space. But as already argued above, there is
no physical motivation for such a prescription.

Given the freedom to choose different regulators for different contact interactions,
it is reasonable to expect that performing more than one subtraction at NLO, i.e.~adding more contact terms, will not qualitatively change the situation.
Since cutoffs for different contact terms are independent, there will
(most probably) exist  ``exceptional'' combinations of them, which
would prevent the existence of a $\Lambda\to\infty$ limit for the amplitude.
The analysis of Sec.~\ref{Sec:LongYang} offers an
example of such a situation  for the case of two NLO contact terms.

\subsection{The case of a local regulator}
\label{Sec:local}
For the sake of completeness, we also briefly comment on the case when
all regulators
are chosen to be local as done e.g.~in Refs.~\cite{Gezerlis:2013ipa,Gezerlis:2014zia,Piarulli:2016vel,Piarulli:2019cqu,Piarulli:2021ywl}
(of course, we could also use a combination of local and non-local regulators which, however,
would not lead to any new conclusions).
In this case, the problem of regularization can be equivalently
formulated in coordinate space
\cite{Valderrama:2009ei,Valderrama:2011mv}.
We adopt the same renormalization conditions as in the previous subsection,
which are specified in Eq.~\eqref{Eq:conditions_p_0},
and  consider again the situation when one subtraction is sufficient
to regularize the NLO amplitude, i.e.~the model with a modified $2\pi$-exchange potential
defined in Sec.~\ref{Sec:formalism}.

There is a  subtlety appearing already at leading order.
In contrast to the previously analyzed case of the non-local regulator,
the solution for the LO constant $C_0(\Lambda_0)$ determined
by the renormalization condition~\eqref{Eq:conditions_p_0}
becomes not unique and depends on the 
number of spurious bound states, see e.g.~Refs.~\cite{Beane:2000wh,Bawin:2003dm,Braaten:2004pg},
leading to different branches of the function $C_0(\Lambda_0)$.
Although the description of the $^3P_0$ partial wave
does not depend on a choice of a particular branch, 
one cannot choose $C_0(\Lambda_0)$ too large because this would affect
other partial waves, in which the $C_0(\Lambda_0)$ contributions are otherwise suppressed by inverse powers of $\Lambda_0$.
To compensate for this effect, one would have to introduce additional 
contact interactions in the affected partial waves.

The NLO amplitude can be conveniently evaluated in $r$-space:
\begin{align}
T^{\rm \scriptscriptstyle NLO{}}(p_\text{on})=\int r^2dr \psi_{p_\text{on}}^{\rm \scriptscriptstyle LO}(r)^2 V^{\rm \scriptscriptstyle NLO{}}(r)=
e^{2i\delta^{\rm \scriptscriptstyle LO}(p_\text{on})}\int r^2dr |\psi_{p_\text{on}}^{\rm \scriptscriptstyle LO}(r)|^2 V^{\rm \scriptscriptstyle NLO{}}(r),
\end{align}
where $\psi_{p_\text{on}}^{\rm \scriptscriptstyle LO}(r)$ is the LO scattering wave function.
Analogously, the contact part of the NLO amplitude is given by
\begin{align}
 T_{\text{ct},0}(p_\text{on})=
e^{2i\delta^{\rm \scriptscriptstyle LO}(p_\text{on})}\int r^2dr |\psi_{p_\text{on}}^{\rm \scriptscriptstyle LO}(r)|^2 V^{\rm \scriptscriptstyle NLO{}}_{\text{ct},0,\Lambda_2}(r).
\label{Eq:T_ct_local}
\end{align}
One can see that the appearance of ``exceptional'' cutoffs 
corresponding to the condition $ T_{\text{ct},0}(p_\text{on})=0$ can be avoided
independently of the behaviour of the LO wave function
if the regulated contact NLO potential does not change its sign.
Obviously, this condition cannot be fulfilled in general.
For example, for the NLO contact interaction in the form of
Eq.~\eqref{Eq:contact_terms_local} with the regulator $\Phi_{\Lambda_2}(q^2)$
employed universally for all four structures, one can show that the coordinate-space
contact potential has the form
(valid only for the $^3P_0$ partial wave)
 \begin{align}
 V^{\rm \scriptscriptstyle NLO{}}_{\text{ct},0,{\Lambda_2}}(r)={\frac{1}{8\pi}}\bigg[ \partial^2_r \Phi_{\Lambda_2}(r)
 -\frac{2}{r}\partial_r \Phi_{\Lambda_2}(r)\bigg],
 \end{align}
with
 \begin{align}
 \Phi_{\Lambda_2} (r)&=\int \frac{d^3 q}{(2\pi)^3}\Phi_{\Lambda_2}(q^2)e^{i\vec q\cdot\vec r}.
 \end{align}
For the power-like regulators of Eq.~\eqref{Eq:local_regulator2} with 
$n=2$ and $n=3$, we obtain
\begin{align}
 V^{\rm \scriptscriptstyle NLO{}}_{\text{ct},0,\Lambda_2}(r)&=\frac{\Lambda_2^4 }{64\pi^2}(2+\Lambda_2  r)\frac{ e^{-\Lambda_2  r}}{r},
 \qquad n=2,\nonumber\\
 V^{\rm \scriptscriptstyle NLO{}}_{\text{ct},0,\Lambda_2}(r)&=\frac{\Lambda_2^5}{256\pi^2} (1+\Lambda_2 r) e^{-\Lambda_2  r},\qquad n=3.
\end{align}
For the Gaussian regulator $\Phi_{\Lambda_2}(q^2)=e^{-q^2/\Lambda_2^2}$,
the corresponding short-range interaction is given by
\begin{align}
 V^{\rm \scriptscriptstyle NLO{}}_{\text{ct},0,\Lambda_2}(r)&=\frac{\Lambda_2^5 }{4\pi^{5/2}}\big(2+\Lambda_2^2  r^2\big)e^{-\Lambda_2^2  r^2}.
\end{align}
None of the above terms changes its sign.
On the other hand, for the regulator in $r$-space in the form
$\Phi_{\Lambda_2}(r)=\Lambda_2^3/[\pi \Gamma(3/4)] e^{-\Lambda_2^4 r^4}$
adopted in Ref.~\cite{Gezerlis:2014zia}, we obtain
\begin{align}
 V^{\rm \scriptscriptstyle NLO{}}_{\text{ct},0,\Lambda_2}(r)&=\frac{\Lambda_2^7 }{2\pi^2 \Gamma(3/4)}r^2
 \big(4\Lambda_2^4  r^4-1\big)e^{-\Lambda_2^4  r^4}.
\end{align}
The latter contact interaction changes its sign at short distances, so that one can tune 
the LO regulator $\Lambda_0$ to make the integral in Eq.~\eqref{Eq:T_ct_local} vanish.
One will also obtain an oscillating behaviour of $V^{\rm \scriptscriptstyle NLO{}}_{\text{ct},0,\Lambda_2}(r)$
if one chooses different cutoff values for different structures 
in Eq.~\eqref{Eq:structures} or if one 
takes a linear combination of the above regulators.

We refrain from providing any numerical results here, which
would essentially duplicate the ones from the previous subsection.
Nevertheless, we have verified explicitly 
the existence of ``exceptional'' cutoffs for the following choice of
the regulator:
\begin{align}
\Phi_{\Lambda_2}(q^2)=2\left( \frac{ \Lambda^2}{\Lambda^2+q^2} \right)^2 
-\left( \frac{\Lambda^2}{\Lambda^2+q^2} \right)^3.
\end{align}

Note that similarly to our comment on the LO interaction,
the constant $C^{\rm \scriptscriptstyle NLO{}}_{0}(\Lambda)$, being infinite for the ``exceptional''
cutoffs, affects also other partial waves even though such contributions are suppressed
by inverse powers of $\Lambda$.

If one includes more than one contact term at NLO,
the conclusion about the existence of ``exceptional'' cutoffs remains the same since such
terms are even more oscillating at short distances, so that the
condition in Eq.~\eqref{Eq:condition_determinant} is likely to be satisfied for certain values of $\Lambda$.

There is a particular choice of $\Lambda_2$ that ensures
the absence of ``exceptional'' cutoff values, namely $\Lambda_2\gg\Lambda_0$.
In this case, the integral in Eq.~\eqref{Eq:T_ct_local} is dominated
by the region $r\sim 1/\Lambda_2$, where the LO wave function 
is not oscillating anymore and approaches its short-range limit (for the $P$-wave) \cite{Newton:1982qc}:
\begin{align}
 \psi_{p_\text{on}}^{\rm \scriptscriptstyle LO}(r)\sim \frac{p_\text{on} r}{f(p_\text{on})}.
\end{align}
Here, $f(p_\text{on})$ is the reduced  Jost function, which is finite at $p_\text{on}=0$.
Since the $P$-wave contact interaction behaves effectively as
\begin{align}
 &V^{\rm \scriptscriptstyle
   NLO{}}_{\text{ct},0,\Lambda_2\to\infty}(r)\sim\partial^2_r\delta(\vec
   r \, )
\end{align}
for $\Lambda_2\to\infty$, the contact term in Eq.~\eqref{Eq:T_ct_local} becomes
\begin{align}
 T_{\text{ct},0}(p_\text{on})/p_\text{on}^2\sim\frac{1}{f(p_\text{on})^2}\ne 0,
\end{align}
and the condition for ``exceptional'' cutoffs is never fulfilled.
The condition $\Lambda_2\gg\Lambda_0$ is quite different from the 
constraint $\Lambda_2=\Lambda_0$ for the absence of ``exceptional'' cutoffs
in the case of non-local regulators, see Sec.~\ref{Sec:nonlocal}.
This indicates, once again, that conditions of this kind have no physical origin.

To summarize, we have argued in this section that the two basic principles of
the infinite-cutoff scheme or, equivalently,  the ``RG-invariant'' EFT framework
stating that
\begin{itemize}
 \item the amplitude has a well defined $\Lambda\to\infty$ limit at
   each EFT order and
 \item this limit does not depend on a particular way it is approached, i.e. on 
 the functional form of the regulators and/or on the relationship among cutoff values 
 at various orders
\end{itemize}
are in conflict with one another.
In general, there is an infinite number of unbounded ``exceptional'' values of the cutoff,
which makes it impossible to formulate a strict infinite-cutoff limit.
We have not proved this rigorously, but we have found several ``exceptional'' cutoffs numerically
in each considered case (apart from some very specific choices of regulators) for rather large cutoff values.

\section{The approach of Long and Yang}
\label{Sec:LongYang}

We are now in the position to examine the results of Ref.~\cite{Long:2011qx}
for the $^3P_0$ NN partial wave at NLO of chiral EFT with respect to
the issues discussed above. 
We do not consider the part of Ref.~\cite{Long:2011qx}
devoted to the N$^2$LO amplitude, which would lead to essentially the same conclusions.
The scheme of Ref.~\cite{Long:2011qx} is very similar to the simplified model with a non-local regulator
considered in Sec.~\ref{Sec:nonlocal}.
However, the two-pion-exchange potential is taken in its full form
given in Eq.~\eqref{Eq:2pi_exchange}, which
requires two subtractions at NLO. The NLO potential is then given by
\begin{align}
V^{\rm \scriptscriptstyle NLO{}}(p',p)=V_{2\pi,\Lambda}(p',p)+ 
p'p \left[C^{\rm \scriptscriptstyle NLO{}}_{0}+C^{\rm \scriptscriptstyle NLO{}}_{2}(p'^2+p^2)\right]F_\Lambda(p')F_\Lambda(p).
\end{align}
Following Ref.~\cite{Long:2011qx}, we implement the sharp cutoff
instead of a smooth non-local regulator,
\begin{align}
F_\Lambda(p)=\theta(\Lambda-p),
\end{align}
with $\Lambda_0=\Lambda_2=\Lambda$,
which makes no conceptual difference to the choices considered before
(even though one would hardly use a sharp cutoff in realistic 
EFT calculations).

The contact parts of the on-shell NLO amplitude can be represented as follows:
\begin{align}
 T_{\text{ct}}(p_\text{on})&=C^{\rm \scriptscriptstyle NLO{}}_{0}T_{\text{ct},0}(p_\text{on})
 +C^{\rm \scriptscriptstyle NLO{}}_{2}T_{\text{ct},2}(p_\text{on}),\nonumber\\
T_{\text{ct},0}(p_\text{on})&=\psi_\Lambda(p_\text{on})^2,\nonumber\\
T_{\text{ct},2}(p_\text{on})&=2\psi_\Lambda(p_\text{on}) \psi'_\Lambda(p_\text{on}),
\label{Eq:contact_parts_Long_Yang}
\end{align}
with
\begin{align}
 \psi_\Lambda(p_\text{on})& =p_\text{on} +\int_0^\Lambda \frac{p^2 dp}{(2\pi)^3}p G(p;p_\text{on})
 T^{\rm \scriptscriptstyle LO}_{\Lambda}(p, p_\text{on};p_\text{on}),\nonumber\\
 \psi'_\Lambda(p_\text{on})& =p_\text{on}^3 +\int_0^\Lambda \frac{p^2 dp}{(2\pi)^3}p^3 G(p;p_\text{on})
 T^{\rm \scriptscriptstyle LO}_{\Lambda}(p, p_\text{on};p_\text{on}).
 \end{align}
The authors of Ref.~\cite{Long:2011qx} adopted the following renormalization conditions
to fix the constants $C^{\rm \scriptscriptstyle LO}_{0}$, $C^{\rm \scriptscriptstyle NLO{}}_{0}$ and $C^{\rm \scriptscriptstyle NLO{}}_{2}$:
\begin{align}
\delta^{\rm \scriptscriptstyle LO}(p_0)&=\delta_\text{exp}(p_0),\nonumber\\
  \delta^{\rm \scriptscriptstyle NLO{}}(p_0)&=0,\nonumber\\
  \delta^{\rm \scriptscriptstyle NLO{}}(p_1)&= \delta_\text{exp}(p_1)-\delta^{\rm \scriptscriptstyle LO}(p_1),
\end{align}
where the on-shell momentum $p_0$ ($p_1$) was chosen to correspond 
to the laboratory energy of $T_\text{lab}=50$~MeV ($T_\text{lab}=25$~MeV).
Using the unitarization prescription in Eqs.~\eqref{Eq:delta2_definition} and~\eqref{Eq:delta2pi_ct_definition}
we obtain the system of linear equations:
\begin{align}
\left\{\begin{array}{ll}
C^{\rm \scriptscriptstyle NLO{}}_{0}\delta_{\text{ct},0}(p_0)+C^{\rm \scriptscriptstyle NLO{}}_{2}\delta_{\text{ct},2}(p_0)&=-\delta_{2\pi}(p_0),\\
C^{\rm \scriptscriptstyle NLO{}}_{0}\delta_{\text{ct},0}(p_1)+C^{\rm \scriptscriptstyle NLO{}}_{2}\delta_{\text{ct},2}(p_1)&=\delta_{\text{exp} }(p_1)-\delta^{\rm \scriptscriptstyle LO}(p_1)-\delta_{2\pi}(p_1).\end{array}\right.
\end{align}
We can express $\delta_{\text{ct},0}$ and $\delta_{\text{ct},2}$ 
in terms of the magnitudes of the vertex functions $\tilde \psi$ and
$\tilde \psi'$,
 \begin{align}
 \psi_\Lambda(p_\text{on})&=\tilde \psi_\Lambda(p_\text{on}) e^{i\delta^{\rm \scriptscriptstyle LO}(p_\text{on})},\nonumber\\
 \psi_\Lambda'(p_\text{on})&=\tilde \psi_\Lambda'(p_\text{on}) e^{i\delta^{\rm \scriptscriptstyle LO}(p_\text{on})},
 \label{Eq:psi_tilde}
 \end{align}
as follows:
\begin{align}
 \delta_{\text{ct},0}(p_\text{on})&=-\tilde \psi_\Lambda(p_\text{on})^2/\rho(p_\text{on}),\nonumber\\
 \delta_{\text{ct},2}(p_\text{on})&=-2\tilde \psi_\Lambda(p_\text{on})\tilde \psi_\Lambda'(p_\text{on})/\rho(p_\text{on}).
\end{align}
Then, the condition for an ``exceptional'' cutoff in Eq.~\eqref{Eq:condition_determinant} becomes
\begin{align}
\det A_{\bar\Lambda}= 2\frac{\tilde\psi_{\bar\Lambda}(p_0)\tilde\psi_{\bar\Lambda}(p_1)}{\rho(p_0)\rho(p_1)}
 \left|\begin{array}{ll}
\tilde\psi_{\bar\Lambda}(p_0)&\tilde\psi'_{\bar\Lambda}(p_0)\\
\tilde\psi_{\bar\Lambda}(p_1)&\tilde\psi'_{\bar\Lambda}(p_1)
\end{array}\right|=0.
\label{Eq:determinant_p0_p1}
\end{align}

\begin{figure}[tb]
\includegraphics[width=\textwidth/2]{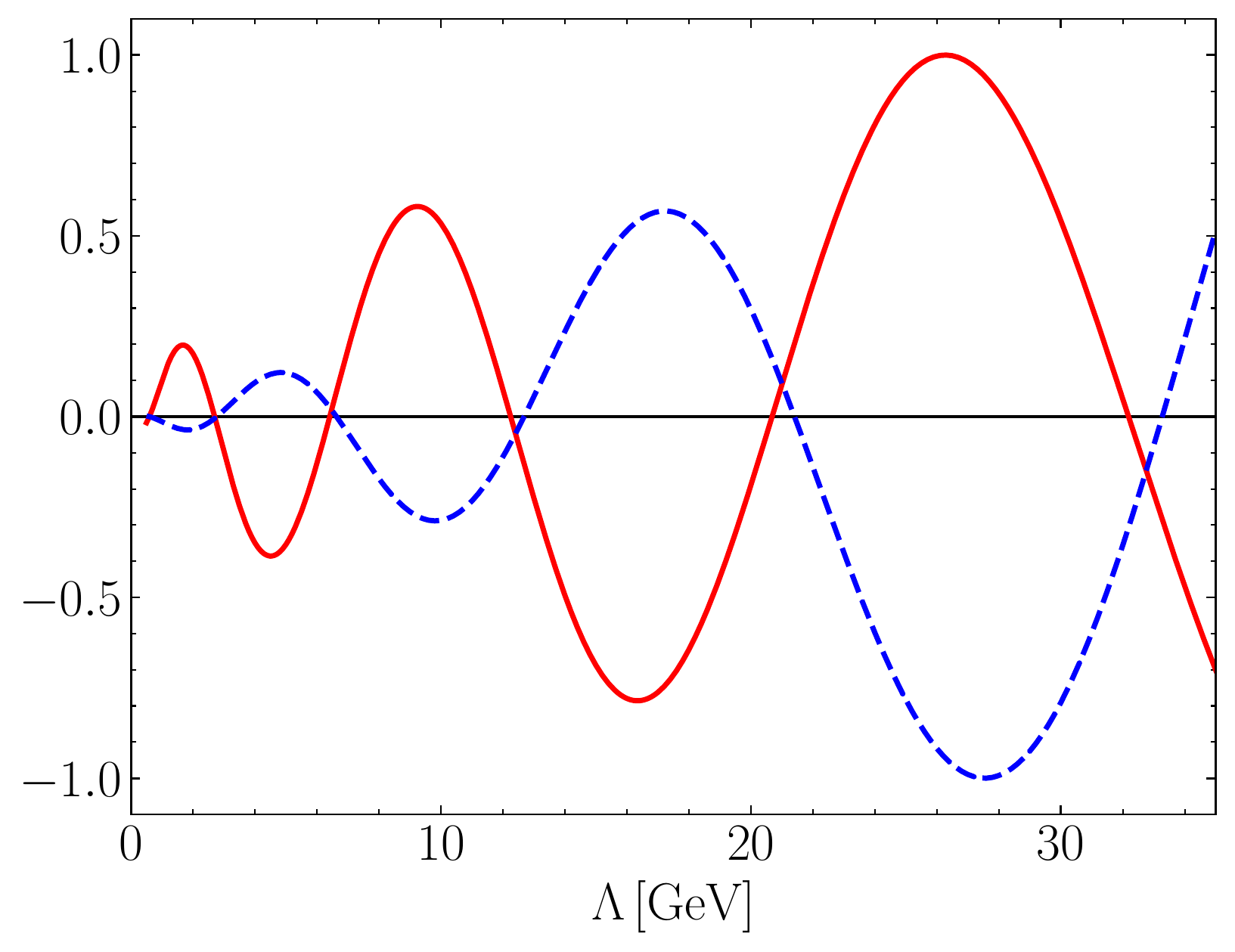}
\caption{$\Lambda$ dependence of the magnitude of the vertex function (dashed line)
$\tilde\psi_{\Lambda}(p_0)$ and of the quantity $\zeta_\Lambda(p_0,p_1)$ (solid line)
for the interaction of Ref.~\cite{Long:2011qx}
normalized to their maximal (in absolute value) values.} \label{Fig:LongYang_determinant}
\end{figure}

As shown in Sec.~\ref{Sec:nonlocal}, the zeros of $\tilde\psi_{\Lambda}(p_0)$
are factorizable, so that $\tilde\psi_{\Lambda}(p_0)=0$ implies $\tilde\psi_{\Lambda}(p_1)=0$.
To exclude these zeros from the determinant in Eq.~\eqref{Eq:determinant_p0_p1}
we introduce the following auxiliary quantity:
\begin{align}
\zeta_\Lambda(p_0,p_1)=\frac{1}{\tilde\psi_\Lambda(p_0)}  \left|\begin{array}{ll}
\tilde\psi_\Lambda(p_0)&\tilde\psi_\Lambda'(p_0)\\
\tilde\psi_\Lambda(p_1)&\tilde\psi'_\Lambda(p_1)
\end{array}\right|,
\end{align}
whose zeros determine the genuine ``exceptional'' cutoffs $\bar\Lambda$:
\begin{align}
\zeta_{\bar\Lambda}(p_0,p_1)=0. 
\end{align}
The quantity $\zeta_\Lambda(p_0,p_1)$ and the magnitude of the vertex function
$\tilde\psi_{\Lambda}(p_0)$ are shown in Fig.~\ref{Fig:LongYang_determinant}
as functions of the cutoff $\Lambda$. They are normalized at the maximal points.
The zeros of $\zeta_\Lambda(p_0,p_1)$ are not factorizable, i.e.~their positions depend on $p_0$ and $p_1$,
and their locations 
(including the lowest ones)
do not coincide with the zeros of $\tilde\psi_{\Lambda}(p_0)$.
Therefore, these zeros indeed correspond to the ``exceptional'' cutoff
values.
Notice that the difference in the positions of zeros of $\zeta_\Lambda(p_0,p_1)$
and $\tilde\psi_{\Lambda}(p_0)$ is an indication of the fact that
the expected connection between ``exceptional'' cutoffs and 
spurious bound states is not direct: the ``exceptional'' cutoffs 
appear at smaller values of $\Lambda$ than the spurious deeply bound states.

\begin{figure}[tb]
\includegraphics{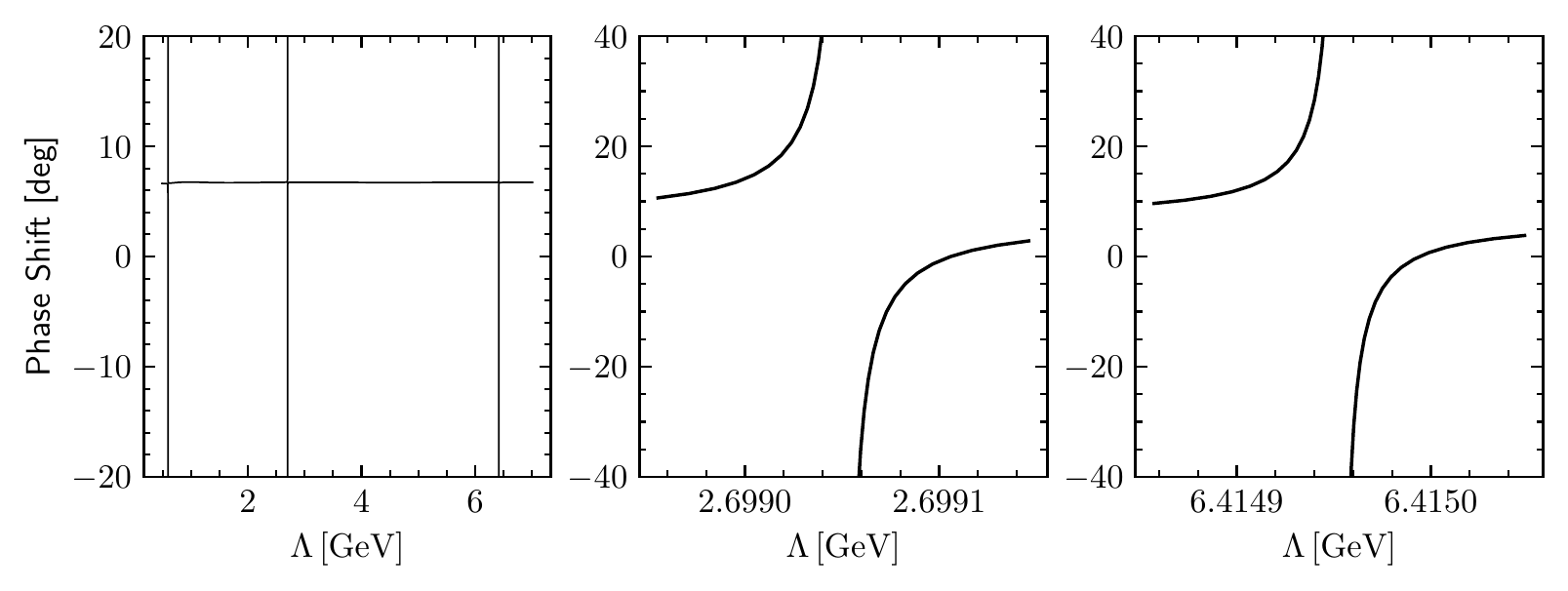}
\caption{Cutoff dependence of $^3P_0$ phase shift calculated at the fixed
  laboratory energy of $T_\text{lab}=130$~MeV 
using the approach of Ref.~\cite{Long:2011qx} at NLO.
The middle and right panels show zoomed regions in the vicinity of two
``exceptional'' cutoffs.} \label{Fig:LongYang_phase_shift_130}
\end{figure}

In the left panel of Fig.~\ref{Fig:LongYang_phase_shift_130}, we plot
the $^3P_0$ phase shift calculated
at NLO at $T_\text{lab}=130$~MeV as a function of the cutoff $\Lambda$.
The curve looks very similar to the one shown in Ref.~\cite{Long:2011qx} 
\footnote{We found the resulting phase shifts to be
  rather sensitive to the employed parameters of the long-range interaction and
  to isospin breaking effects in the one-pion-exchange potential.}
and seems to flatten out as $\Lambda$ tends to infinity.
The only differences to the plots in Ref.~\cite{Long:2011qx} are the
almost vertical
lines located at the ``exceptional'' values of the cutoff.
The middle and right panels of Fig.~\ref{Fig:LongYang_phase_shift_130}
demonstrate the behaviour of the phase shift around two ``exceptional''
points in more detail. 
As one can see, the width of such ``exceptional'' regions is
of order $0.1$~MeV, which is much smaller than in the case of 
the simplified model with one subtraction considered in Sec.~\ref{Sec:nonlocal}.
The reason for this is obviously the fact that two subtractions
performed at NLO lead to a higher power $\beta$ of the residual 
cutoff dependence in Eq.~\eqref{Eq:T2_Lambda}.
In Ref.~\cite{Long:2011qx}, an estimate $\beta=5/2$ was given
based on the short-range behaviour of the LO wave function.
Another factor that makes the ``exceptional'' regions
narrower is that the zeros of $\zeta_{\bar\Lambda}(p_0,p_1)$ are first order,
see Eq.~\eqref{Eq:deltaLambda}.

\begin{figure}[tb]
 \includegraphics{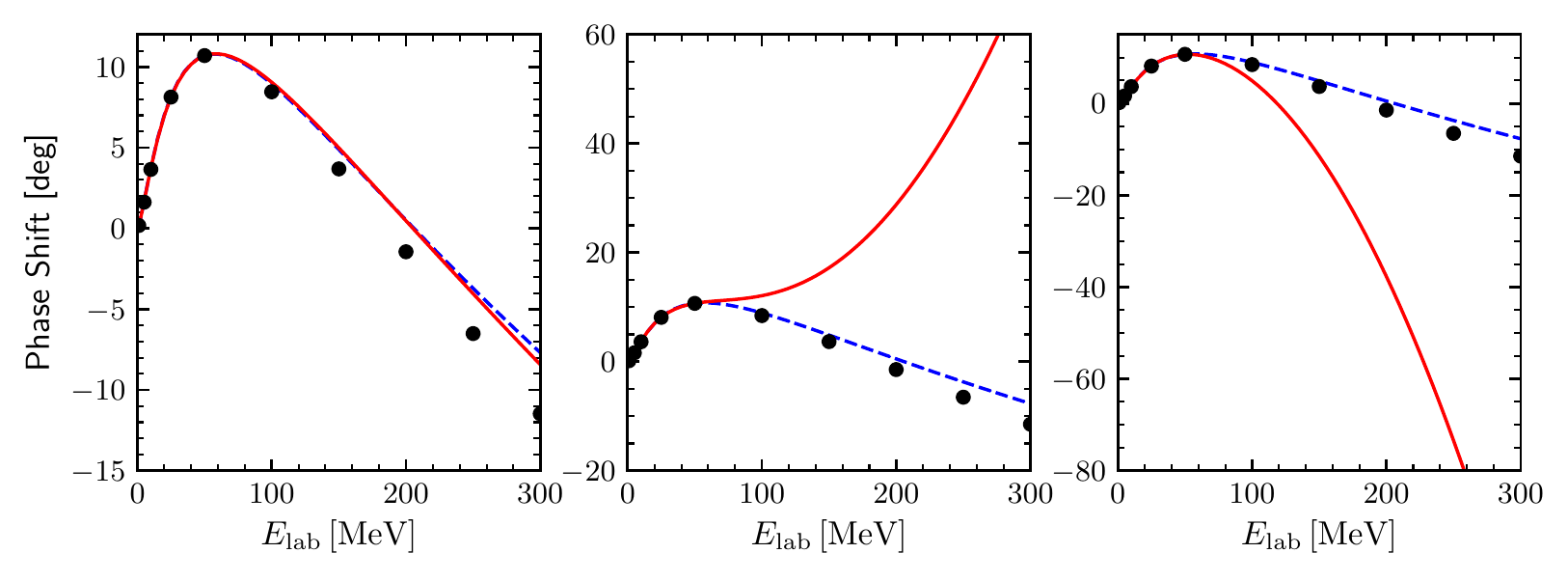}
 \caption{The $^3P_0$ phase shifts calculated using the approach of Ref.~\cite{Long:2011qx} at
LO (dashed lines) and NLO (solid lines).
The left panel shows the solution for a typical cutoff ($\Lambda=12.5$~GeV),
whereas the middle and right panels  
correspond to the cutoff values slightly below ($\Lambda=12249.69$~MeV) and above ($\Lambda=12249.73$~MeV)
the second ``exceptional'' point.
 } \label{Fig:LongYang_phase_shifts}
\end{figure}

The difference between the behaviour of the phase shifts for
the typical and ``exceptional'' values of the cutoff is illustrated
in Fig.~\ref{Fig:LongYang_phase_shifts}.
The left panel shows the solution for a typical cutoff ($\Lambda=12.5$~GeV),
whereas the middle and the right panels  
correspond to cutoff values slightly below ($\Lambda=12249.69$~MeV) and above ($\Lambda=12249.73$~MeV)
an ``exceptional'' point. This results demonstrate that the NLO phase
shifts show essentially arbitrary behavior for cutoffs in the vicinity
of the ``exceptional'' values, which signals the breakdown of the
renormalization program.   

It is important to keep in mind that the
appearance of an infinite number of ``exceptional'' cutoffs does not depend on the choice of
renormalization conditions, which only determine the particular
locations of such cutoffs. Moreover, adding further counter terms and fixing
them using additional renormalization conditions would not qualitatively
change the situation because Eq.~\eqref{Eq:condition_determinant}
would still have solutions, albeit the width of ``exceptional''
regions would further decrease.

To summarize, the results presented above confirm the analysis based on the simplified model of Sec.~\ref{Sec:nonlocal}.
The realistic calculation for the NN system indicates the absence of a definite limit 
of the NLO amplitude for $\Lambda\to\infty$.
We have also highlighted the danger of missing the ``exceptional''
cutoff regions when numerically verifying the cutoff
independence of an amplitude, since their width may be very small.

\section{Generalizations}
\label{Sec:Generalizations}
We are now in the position to discuss various straightforward generalizations
of the results obtained in Secs.~\ref{Sec:general}~and~\ref{Sec:LongYang}.
Since the existence of ``exceptional'' cutoffs originates from the
oscillating nature of the high-momentum part of the $T$-matrix for singular potentials,
such cutoffs are expected to generally appear in any infinite-cutoff scheme with the 
non-perturbative LO attractive singular potential and perturbative
inclusion of the NLO corrections. This implies, in particular, that
the renormalizability in the sense of the ``RG-invariant'' EFT
framework breaks down in all spin-triplet channels where the
one-pion-exchange potential is attractive and needs to be treated 
non-perturbatively.\footnote{In fact, the existence of even a single
  problematic partial wave would already put the ``RG-invariant'' EFT
  approach in question.}  

Moreover, the obtained results do not qualitatively depend on the long-range
part of the LO potential. In particular, we have verified that the  
``exceptional'' cutoffs exist also in the case when the
one-pion-exchange potential is taken in the chiral limit
with $M_\pi=0$, as suggested in Ref.~\cite{Beane:2001bc}.

It is easy to check that ``exceptional'' cutoffs appear also in systems where
the LO interaction is less singular, e.g.~where the LO
potential at short distances behaves as
$1/r^2$ and  is sufficiently attractive (one exception being the model of Long and van Kolck~\cite{Long:2007vp}, 
which will be discussed in Sec.~\ref{Sec:LongVanKolck}).
The issues raised in our paper are therefore also relevant for
studies of three-body systems with zero-range interactions,
such as e.g.~three-nucleon scattering 
in the doublet $S$-wave channel treated within pionless EFT~\cite{Bedaque:1999ve}
or three-boson systems with large scattering lengths~\cite{Bedaque:1998km}.
In those cases, the leading-order interaction with an infinite cutoff 
leads to the Skorniakov-Ter-Martirosian equation~\cite{Skornyakov1957},
which in the ultraviolet regime is essentially equivalent 
to the Lippmann-Schwinger equation with a $1/r^2$-potential.
The inclusion of higher-order corrections in perturbation theory
as done e.g.~in Refs.~\cite{Hammer:2001gh,Vanasse:2013sda,Ji:2011qg,Ji:2012nj}
inevitably leads to the issues discussed in the previous sections.

Finally, it is clear that the same issues will also be relevant in applications beyond the NN scattering problem,
especially where the results are sensitive to the short-range part of the NN wave function,
e.g.~for  few- (many-) nucleon systems or electroweak processes involving
several nucleons.

\section{The toy model of Long and Van Kolck}
\label{Sec:LongVanKolck}

As already pointed out above, one notable exception of the general
situation with the appearance of ``exceptional'' cutoffs is the toy
model introduced by Long and van Kolck \cite{Long:2007vp}.
The model is particularly interesting since many of the results can be
derived analytically in a closed form. Below, we analyze the peculiar
features of this model that result in the absence of genuine
``exceptional'' cutoffs. We also argue that a slight modification of
the LO interaction in this model and/or the employed regulator leads
to the breakdown of the renormalization program at NLO.  

The model of Long and van Kolck has a lot of similarities with the scheme considered in Sec.~\ref{Sec:LongYang}.
We consider $S$-wave two-body scattering of particles with the
mass $m_N$ based on the LO and NLO potentials
$V^{\rm \scriptscriptstyle LO}$ and $V^{\rm \scriptscriptstyle NLO}$
given by a sum of the long-range and short-range terms,
\begin{align}
 V^{\rm \scriptscriptstyle LO}&=V_{\text{L}}^{\rm \scriptscriptstyle LO}+V_\text{S}^{\rm \scriptscriptstyle LO},\qquad
 V^{\rm \scriptscriptstyle NLO}=V_{\text{L}}^{\rm \scriptscriptstyle NLO}+V_\text{S}^{\rm \scriptscriptstyle NLO}.
\end{align}
The long-range part of the LO potential
has the form
\begin{align}
 V_{\text{L}}^{\rm \scriptscriptstyle LO}(p', p) &= - \frac{8\pi^3 \lambda}{m_N}\, \frac{1}{p_>}.
\end{align}
Here and below, the notation $p_>=\max(p',p)$ and $p_<=\min(p',p)$ is used.
The potential $V_{\text{L}}^{\rm \scriptscriptstyle LO}$ is proportional to the Fourier transform
of the function $1/r^2$.
The coupling constant $\lambda$ is chosen 
such that it corresponds to a singular LO potential
($\lambda=2$ in Ref.~\cite{Long:2007vp}).

The long-range part of the NLO potential is given by
\begin{align}
V_{\text{L}}^{\rm \scriptscriptstyle NLO}(p', p) &= 
\sum_{i=0}^3 g_i V_{\text{L},i}^{\rm \scriptscriptstyle NLO}(p', p),
\label{Eq:V_L_NLO_general_a}
\end{align}
with
\begin{align}
V_{\text{L},0}^{\rm \scriptscriptstyle NLO}(p', p) &= \frac{8\pi^3  M_\pi^2 }{m_N^3}\, \frac{1}{p_>},\nonumber\\
V_{\text{L},1}^{\rm \scriptscriptstyle NLO}(p', p) &=\frac{8\pi^3 }{m_N^3} p_>,\nonumber\\
V_{\text{L},2}^{\rm \scriptscriptstyle NLO}(p', p) &=\frac{8\pi^3 }{m_N^3} \frac{p_<^2}{p_>}.
\label{Eq:V_L_NLO_general_b}
\end{align}
In Ref.~\cite{Long:2007vp} the following choice for the coupling constants is made:
$g_0=0$, $g_2=g_1/3$.
In such a case, $V_{\text{L}}^{\rm \scriptscriptstyle NLO}$
is proportional to the Fourier transform
of the function $1/r^4$ up to a contact interaction.
We adopt a more general ansatz for $V_{\text{L}}^{\rm \scriptscriptstyle NLO}$
as will turn out to be useful in the subsequent analysis.
The short-range parts of the LO and NLO potentials are given by
\begin{align}
 V_\text{S}^{\rm \scriptscriptstyle LO}(p', p) &= C_0(\Lambda),\nonumber\\
  V_\text{S}^{\rm \scriptscriptstyle NLO}(p', p) &= C_0^{\rm \scriptscriptstyle NLO}(\Lambda)+C_2^{\rm \scriptscriptstyle NLO}(\Lambda)(p'^2+p^2).
\end{align}
Finally, we follow Refs.~\cite{Long:2011qx,Long:2007vp} and employ the sharp regulator
with the cutoff $\Lambda$. 
For the sake of convenience, we reshuffle the regulator
from the potential to the propagator by introducing
\begin{equation}
G_\Lambda(p; p_\text{on})=\frac{m_N}{p_\text{on}^2-p^2+i \epsilon}\theta(\Lambda-p).
\end{equation}  

The LO amplitude $T^{\rm \scriptscriptstyle LO}$ is obtained by solving the Lippmann-Schwinger equation (see Eq.~\eqref{Eq:LS_equation}):
\begin{align}
T^{\rm \scriptscriptstyle LO}=V^{\rm \scriptscriptstyle LO}+V^{\rm \scriptscriptstyle LO}G_\Lambda T^{\rm \scriptscriptstyle LO},
\label{Eq:LS_LongKolck}
\end{align}
whereas the NLO amplitude is given in terms of the distorted-wave Born approximation:
\begin{align}
T^{\rm \scriptscriptstyle NLO}=(\mathds{1}+T^{\rm \scriptscriptstyle LO}G_\Lambda) V^{\rm \scriptscriptstyle NLO} (\mathds{1}+G_\Lambda T^{\rm \scriptscriptstyle LO}).
\end{align}
The on-shell NLO amplitude can be represented, by analogy with Eq.~\eqref{Eq:contact_parts_Long_Yang},
via
\begin{align}
T^{\rm \scriptscriptstyle NLO}(p_\text{on}) =T_{\text{L}}^{\rm \scriptscriptstyle NLO}(p_\text{on}) 
+  C_0^{\rm \scriptscriptstyle NLO}(\Lambda)\psi_\Lambda(p_\text{on})^2 
+ 2 C_2^{\rm \scriptscriptstyle NLO}(\Lambda) \psi_\Lambda(p_\text{on}) \psi'_\Lambda(p_\text{on})
\end{align}
with
\begin{align}
 T_{\text{L}}^{\rm \scriptscriptstyle NLO}=(\mathds{1}+T^{\rm \scriptscriptstyle LO}G_\Lambda) V_{\text{L}}^{\rm \scriptscriptstyle NLO} (\mathds{1}+G_\Lambda T^{\rm \scriptscriptstyle LO}),
\end{align}
and
\begin{align}
 \psi_\Lambda(p;p_\text{on})& =1 +\int \frac{p'^2 dp'}{(2\pi)^3} G_\Lambda(p';p_\text{on})
 T^{\rm \scriptscriptstyle LO}(p', p;p_\text{on}),\nonumber\\
 \psi_\Lambda(p_\text{on})& \coloneqq\psi_\Lambda(p_\text{on};p_\text{on}),\nonumber\\
 \psi'_\Lambda(p_\text{on})& =p_\text{on}^2 +\int \frac{p^2 dp}{(2\pi)^3}p^2 G_\Lambda(p;p_\text{on})
 T^{\rm \scriptscriptstyle LO}(p, p_\text{on};p_\text{on}).
 \label{Eq:psi_psi_prime}
 \end{align}
Similarly to Ref.~\cite{Long:2011qx}, the authors of Ref.~\cite{Long:2007vp} adopted the following renormalization conditions
to fix the constants $C^{\rm \scriptscriptstyle LO}_{0}$, $C^{\rm
  \scriptscriptstyle NLO{}}_{0}$ and $C^{\rm \scriptscriptstyle
  NLO{}}_{2}$ \footnote{Strictly speaking, these conditions are
  formulated in terms of the $K$-matrix in Ref.~\cite{Long:2007vp}.}:
\begin{align}
\delta^{\rm \scriptscriptstyle LO}(p_0)&=\delta_\text{exp}(p_0),\nonumber\\
  \delta^{\rm \scriptscriptstyle NLO}(p_0)&=0,\nonumber\\
  \delta^{\rm \scriptscriptstyle NLO}(p_1)&= \delta_\text{exp}(p_1)-\delta^{\rm \scriptscriptstyle LO}(p_1),
  \label{Eq:renormalization_condition_LongKolck}
\end{align}
where $p_0$ and $p_1$ are some center-of-mass momenta and $p_0, p_1\ll\Lambda$.
As follows from the analysis of Sec.~\ref{Sec:LongYang},
if there were ``exceptional'' values of the cutoff,
they would correspond to 
the zeros of the quantity
\begin{align}
\zeta_\Lambda(p_0,p_1)=\frac{1}{\tilde\psi_\Lambda(p_0)}  \left|\begin{array}{ll}
\tilde\psi_\Lambda(p_0)&\tilde\psi_\Lambda'(p_0)\\
\tilde\psi_\Lambda(p_1)&\tilde\psi'_\Lambda(p_1)
\end{array}\right|.
\label{Eq:LongVanKolck_determinant}
\end{align}

To proceed further, we first derive the important relationship between
the functions $\psi_\Lambda(p_\text{on})$ and
$\psi'_\Lambda(p_\text{on})$ that holds for the considered model. 
For this, we rewrite $\psi'_\Lambda(p_\text{on})$ as 
\begin{align}
  \psi'_\Lambda(p_\text{on})& =p_\text{on}^2\psi_\Lambda(p_\text{on})
  -m_N\int_0^\Lambda \frac{p^2 dp}{(2\pi)^3}
 T^{\rm \scriptscriptstyle LO}(p, p_\text{on};p_\text{on})\nonumber\\
 &=p_\text{on}^2\psi_\Lambda(p_\text{on})
 +\bigg[\frac{\lambda}{2}\Lambda^2
 -\frac{C_0(\Lambda)}{24\pi^3} m_N\Lambda^3\bigg]\psi_\Lambda(p_\text{on})
 -\frac{\lambda}{6}\psi'_\Lambda(p_\text{on}),
 \label{Eq:equation_psi_prime}
 \end{align}
where the last equality is obtained by performing a single iteration of the 
Lippmann-Schwinger equation for $T^{\rm \scriptscriptstyle LO}$
and calculating explicitly the integral 
 \begin{align}
\int_0^\Lambda \frac{p'^2 dp'}{(2\pi)^3}
 V^{\rm \scriptscriptstyle LO}(p', p)&=
C_0(\Lambda) \int_0^\Lambda \frac{p'^2 dp'}{(2\pi)^3}
 - \frac{8\pi^3 \lambda}{m_N}\int_0^\Lambda \frac{p'^2 dp'}{(2\pi)^3}\frac{1}{\max(p',p)} \nonumber\\
 &=\frac{C_0(\Lambda)}{24\pi^3} \Lambda^3
 -\frac{\lambda}{2m_N}\Lambda^2+\frac{\lambda}{6m_N}p^2.
  \end{align}
Solving Eq.~\eqref{Eq:equation_psi_prime} with respect to
$\psi'_\Lambda$ we obtain the desired relationship in the form
\begin{align}
  \psi'_\Lambda(p_\text{on})& =
  (\gamma_1(\Lambda) \Lambda^2
  +  \gamma_2 p_\text{on}^2)\psi_\Lambda(p_\text{on}),
  \label{Eq:psi_prime_via_psi}
 \end{align}
 where the quantities $\gamma_1(\Lambda) $ and $\gamma_2$ are given by
 \begin{align}
  \gamma_1(\Lambda)&=\frac{6}{6+\lambda} \bigg[\frac{\lambda}{2}
 -\frac{C_0(\Lambda)}{24\pi^3} m_N\Lambda\bigg],\nonumber\\
 \gamma_2&=\frac{6}{6+\lambda}.
 \label{Eq:gamma1_gamma2}
 \end{align} 
Notice that Eq.~\eqref{Eq:psi_prime_via_psi}
holds true exactly and not just approximately up to terms of some order
$O[\left(\bar q/\Lambda\right)^{\alpha_{\text{ct}}}]$, cf.~Eq.~\eqref{E:N_infty}.
Such corrections were responsible for the appearance of
``exceptional'' cutoff values discussed in the previous sections.

It is now easy to see from Eq.~\eqref{Eq:psi_prime_via_psi} that the
quantity $\zeta_\Lambda(p_0,p_1)$, defined in
Eq.~(\ref{Eq:LongVanKolck_determinant}), can be written as 
\begin{align}
\zeta_\Lambda(p_0,p_1)=\gamma_2 (p_1^2-p_0^2)\tilde\psi_\Lambda(p_1).
\label{Eq:LongVanKolck_determinant2}
\end{align}
Therefore, the zeros of $\zeta_\Lambda(p_0,p_1)$ coincide with the zeros
of $\psi_\Lambda(p_1)$ (and of $\psi_\Lambda(p_0)$).
Moreover, they factorize according to Eq.~\eqref{Eq:zeros_factorization}
and correspond to cutoffs $\bar\Lambda$ for which $C_0(\bar\Lambda)=\infty$.

It is, however, important to emphasize that even a slight modification of
the underlying model by adding e.g.~a logarithmic factor similar to the 
one appearing in the Skorniakov-Ter-Martirosian equation
to the LO long-range potential, changing the sharp regulator to a
smooth one or using  different values of the LO and NLO cutoffs
lead to a violation of Eq.~\eqref{Eq:psi_prime_via_psi} and,
therefore, results in the appearance of ``exceptional'' cutoffs that destroy
the renormalizability of the NLO amplitude.

Finally, while the factorization of the zeros of $\zeta_\Lambda(p_0,p_1)$
prevents the existence of ``exceptional'' cutoffs for the considered
toy model, this feature alone does not necessarily prove the cutoff
independence of the NLO amplitude in a strict mathematical sense.
The simplicity of the considered model and the absence of
scales in the long-range parts of the interaction make
it possible to carry out such a  proof analytically.
An attempt to provide the proof was already made in Ref.~\cite{Long:2007vp}.
However, some important steps were missing there, and some
conclusions were not justified.
Moreover, the solution for the
cutoff dependence of the NLO LECs was not provided.
In Appendices \ref{Sec:AppenLO} and  \ref{Sec:AppenNLO}, we fill these gaps by proving the cutoff
independence of the LO and NLO amplitudes in the $\Lambda \to \infty$
limit and deriving the explicit solutions
for $C_0^{\rm \scriptscriptstyle NLO}(\Lambda)$ and $C_2^{\rm
  \scriptscriptstyle NLO}(\Lambda)$.

\section{Summary and conclusions}
\label{Sec:sum}
We have studied two-nucleon scattering using the
chiral EFT framework formulated in
Refs.~\cite{Long:2007vp,Hammer:2019poc,vanKolck:2020llt,
  Long:2011qx,Long:2011xw}, which permits the usage of arbitrarily large
cutoffs and is claimed to be RG-invariant in the $\Lambda \to \infty$ limit.
In this scheme, the one-pion-exchange potential is iterated to all
orders in low partial waves together with the necessary counter terms,
while subleading corrections to the amplitude are taken into account
using the distorted-wave Born approximation. Renormalizability within
this method depends upon the fulfillment of two requirements:
(i) the scattering amplitude should possess
a well-defined limit when the cutoff $\Lambda$ tends to infinity
 at each EFT  order individually and (ii) this limit should not depend
on a particular form of regulator. The existing calculations within this
scheme are summarized in a recent review article by van Kolck \cite{vanKolck:2020llt},
who then concludes that ``the longstanding problem of renormalization of chiral
nuclear forces has been solved at the 2N and 3N levels''.
Even leaving
aside the criticism of the ``RG-invariant'' scheme raised in
Refs.~\cite{Epelbaum:2009sd,Epelbaum:2018zli,Epelbaum:2020maf,Epelbaum:2021sns},
the results of our study show that this conclusion is too optimistic.
Specifically, we have demonstrated that the above-mentioned renormalizability
requirements of the ``RG-invariant'' approach can, in general, not be fulfilled
simultaneously beyond LO. The problem is related to the existence 
of ``exceptional'' cutoff values, in the vicinity of which the 
renormalization of the NLO amplitude breaks down. The 
``exceptional'' $\Lambda$-values extend to infinity and originate from the 
oscillatory short-distance behaviour of the LO wave function caused by  
the singular nature of the one-pion-exchange potential.
While we have specifically focused in this paper on the $^3P_0$ channel of NN
scattering and limited ourselves to NLO, our conclusions apply to all 
attractive spin-triplet channels (in which the one-pion-exchange
potential is iterated to all orders) and remain valid beyond NLO. 
The main results of our study can be summarized as follows:
\begin{itemize}
\item
We have given general qualitative arguments illustrating the
issues with the ``exceptional'' cutoff values 
based on the dispersive representation of the
scattering amplitude. To substantiate these findings we studied
the effects of the ``exceptional'' cutoff values
in a simplified model, where the two-pion-exchange potential was
modified to require only one subtraction (i.e., a single contact term)
at NLO, using a non-local regulator. Our numerical results 
reveal a clear (unbounded) deviation of the $^3P_0$ phase shifts for
cutoff values in the vicinity of the ``exceptional'' ones  
as compared to the typical $\Lambda$-values as depicted in
Fig.~\ref{Fig:ToyModel_phase_shifts}. 
We also argued that including additional contact interactions would lead to
essentially the same conclusions, and it would not restore the renormalizability.
The case of locally regulated interactions has also been discussed
in order to demonstrate the general nature of the considered arguments
and the independence of results on a particular regulator choice.
\item
As the next application, we have examined the calculation of the
$^3P_0$ scattering amplitude by Long and Yang \cite{Long:2011qx} up to NLO in
chiral EFT. We used the same renormalization conditions as done in
Ref.~\cite{Long:2011qx} to fix one
counter term at LO and two additional counter terms at NLO. We observed that ``exceptional'' cutoff values 
occur in this case as well, and they prevent the existence of the
$\Lambda\to\infty$ limit of the scattering amplitude and phase shifts,
see Figs.~\ref{Fig:LongYang_phase_shift_130} and \ref{Fig:LongYang_phase_shifts}.
The corresponding problematic cutoff regions appear to be rather narrow
and can be easily overlooked when performing numerical
checks.
\item
  We have argued that our findings are relevant for a broad class of
  problems studied using similar EFT frameworks. These include, but
  are not limited to, the
  proposal of calculating NN scattering using an expansion of nuclear
  forces about the chiral limit \cite{Beane:2001bc} and applications of pionless
  EFT to study the 3-body problem near the unitary limit using 
  the Skorniakov-Ter-Martirosian equation 
  \cite{Hammer:2001gh,Vanasse:2013sda,Ji:2011qg,Ji:2012nj}.
  Generally, artifacts similar to the ones considered in the present paper
 are expected to appear in applications beyond the two-nucleon system, whenever the short
range part of the LO amplitude plays a significant role. 
\item
  Finally, we have analyzed in detail the renormalization of the
  scattering amplitude for the toy model of Long and van Kolck 
\cite{Long:2007vp}, for which  many results can be derived analytically.
We delivered a rigorous and complete proof of the cutoff independence of the LO and NLO amplitudes
and provided explicit solutions for the $\Lambda$ dependence of the NLO LECs.
The absence of genuine ``exceptional'' cutoff values is a peculiar
feature of this model, which is
found to depend crucially on the form of the LO interaction (a pure
$1/r^2$-potential) and on the particular regularization scheme (the
same sharp regulator for the LO and NLO terms). Even a slight
modification of these features of the model is expected to result in the emergence of
the ``exceptional'' cutoff values, which would destroy its
renormalizability beyond LO. 
\end{itemize}

As an alternative to the ``RG-invariant'' approach, few-nucleon
systems are being successfully analyzed within the finite-cutoff
formulation of chiral EFT using $\Lambda \sim \Lambda_b$, see e.g.~\cite{Epelbaum:2019kcf} for a review
article. In this scheme, the amplitude calculated at any finite EFT
order is only approximately cutoff independent, while the  
exact cutoff independence is only achievable upon taking into account the
contributions of an infinite number of counter terms from the
effective Lagrangian. Recently, a rigorous renormalizability proof of this
scheme to NLO, valid to all orders in the iterations of
the LO potential, was accomplished by explicitly demonstrating that all
power-counting breaking terms are absorbable into a redefinition of
the available  LECs \cite{Gasparyan:2021edy}. At the same time, the
method proposed in that paper allows one to systematically eliminate
regulator artifacts from the calculated observables. A generalization
of the renormalizability proof of Ref.~\cite{Gasparyan:2021edy} to
purely non-perturbative channels is in progress.

\section*{Acknowledgments} 
We are grateful to Jambul Gegelia for sharing his insights into
the considered topics. We also thank Jambul Gegelia and
Ulf-G. Mei{\ss}ner for helpful comments on the manuscript.
This work was supported by DFG (Grant No. 426661267), by BMBF (contract No. 05P21PCFP1), by ERC AdG
NuclearTheory (grant No. 885150) and by the EU Horizon 2020 research and
innovation programme (STRONG-2020, grant agreement No. 824093).

\newpage
\appendix
\section{Renormalizability proof for the model of Long and van Kolck:
  LO analysis}
\label{Sec:AppenLO}
We start with proving the renormalizability for the LO amplitude. To
show the cutoff independence of $T^{\rm \scriptscriptstyle LO}$ in the
$\Lambda \to \infty$ limit, we follow Ref.~\cite{Long:2007vp} and differentiate
the Lippmann-Schwinger equation~\eqref{Eq:LS_LongKolck} with respect to $\Lambda$:
\begin{align}
 \frac{d}{d\Lambda}T^{\rm \scriptscriptstyle LO}(p',p;p_\text{on})&=
 \frac{\mathcal{M}(\Lambda)}{(2\pi)^3}\frac{\Lambda^2 m_N}{p_\text{on}^2-\Lambda^2}T^{\rm \scriptscriptstyle LO}(\Lambda,p;p_\text{on}) \nonumber\\
 &+\frac{dC_0(\Lambda)}{d\Lambda}\psi_\Lambda(p;p_\text{on})\nonumber\\
 & +\int \frac{p''^2dp''}{(2\pi)^3}V^{\rm \scriptscriptstyle LO}(p',p'')G_\Lambda(p'';p_\text{on})\frac{d}{d\Lambda}T^{\rm \scriptscriptstyle LO}(p'',p;p_\text{on}),
 \label{Eq:dT0_dLambda}
\end{align}
where we have used that
\begin{align}
 \frac{d}{d\Lambda} V^{\rm \scriptscriptstyle LO}(p',p)=\frac{dC_0(\Lambda)}{d\Lambda},
\end{align}
and
\begin{align}
 V^{\rm \scriptscriptstyle LO}(\Lambda,p)=V^{\rm \scriptscriptstyle LO}(p',\Lambda)=-\frac{(2\pi)^3\lambda}{m_N\Lambda}+C_0(\Lambda)\eqqcolon\mathcal{M}(\Lambda).
 \label{Eq:defifition_Mcal}
\end{align}
The amplitude $T^{\rm \scriptscriptstyle LO}(\Lambda,p;p_\text{on})$
can be written as
\begin{align}
 T^{\rm \scriptscriptstyle LO}(\Lambda,p;p_\text{on})&=\mathcal{M}(\Lambda)\psi_\Lambda(p;p_\text{on}),
 \label{Eq:T_Lambda}
\end{align}
so that Eq.\eqref{Eq:dT0_dLambda} becomes
\begin{align}
 \frac{d}{d\Lambda}T^{\rm \scriptscriptstyle LO}(p',p;p_\text{on})&=
 \left[\frac{dC_0}{d\Lambda}-\frac{\mathcal{M}(\Lambda)^2 m_N}{(2\pi)^3}\right]
 \psi_\Lambda(p;p_\text{on})\nonumber\\
 &-\frac{p_\text{on}^2}{\Lambda^2-p_\text{on}^2}\frac{\mathcal{M}(\Lambda)^2 m_N}{(2\pi)^3}
 \psi_\Lambda(p;p_\text{on})\nonumber\\
  & +\int \frac{p''^2dp''}{(2\pi)^3}V^{\rm \scriptscriptstyle LO}(p',p'')G_\Lambda(p'';p_\text{on})\frac{d}{d\Lambda}T^{\rm \scriptscriptstyle LO}(p'',p;p_\text{on}).
  \label{Eq:dT0_dLambda_2}
\end{align}
Since we are looking for a cutoff-independent solution in the limit $\Lambda\to\infty$,
we neglect  in Eq.~\eqref{Eq:dT0_dLambda_2} all terms involving $\frac{d}{d\Lambda}T^{\rm \scriptscriptstyle LO}$ and 
terms of order $\sim p_\text{on}^2/\Lambda^2$
to obtain the equation for $C_0 (\Lambda)$
\begin{align}
 \frac{dC_0 (\Lambda)}{d\Lambda}-\frac{\mathcal{M}(\Lambda)^2 m_N}{(2\pi)^3}=0,
 \label{Eq:equation_for_C0}
\end{align}
which can be solved explicitly yielding
\begin{equation}
 C_0(\Lambda) = - \frac{(2\pi)^3\lambda}{\Lambda m_N} 
  \frac{1 - 2 \nu \tan [\nu\ln (\Lambda / \Lambda_{*})]}
       {1 + 2 \nu \tan [\nu\ln (\Lambda / \Lambda_{*})]},\qquad
       \lambda=\frac{1}{4}+\nu^2,
       \label{Eq:solution_C0}
\end{equation}
where $\Lambda_{*}$ is determined by the renormalization condition~\eqref{Eq:renormalization_condition_LongKolck}
at $\Lambda\to\infty$.

In this derivation, we assumed that 
neglecting certain terms in Eq.~\eqref{Eq:dT0_dLambda_2} as described above is justified.
However, this is not  obvious for the neglected terms on the right-hand side
of Eq.~\eqref{Eq:dT0_dLambda_2} when the cutoff takes values close to
``exceptional'' ones, $\Lambda\sim\bar\Lambda $ with $
C_0(\bar\Lambda )=\infty$, since the neglected 
terms involve the prefactors $C_0(\Lambda)$ and $C_0(\Lambda)^2$.
To clarify this issue we substitute the solution~\eqref{Eq:solution_C0} for $C_0(\Lambda)$ 
into Eq.~\eqref{Eq:dT0_dLambda_2}:
\begin{align}
\left[(\mathds{1}-V^{\rm \scriptscriptstyle LO}G_\Lambda)\frac{dT^{\rm \scriptscriptstyle LO}}{d\Lambda}\right](p',p;p_\text{on})&=
-\frac{p_\text{on}^2}{\Lambda^2-p_\text{on}^2}\frac{\mathcal{M}(\Lambda)^2 m_N}{(2\pi)^3}\psi_\Lambda(p;p_\text{on}).
\end{align}
The solution to this equation with respect to $\frac{d}{d\Lambda}T^{\rm \scriptscriptstyle LO}$ reads:
\begin{align}
 \frac{d}{d\Lambda}T^{\rm \scriptscriptstyle LO}(p',p;p_\text{on})&=
-\frac{p_\text{on}^2}{\Lambda^2-p_\text{on}^2}\frac{\mathcal{M}(\Lambda)^2 m_N}{(2\pi)^3}
\psi_\Lambda(p';p_\text{on})\psi_\Lambda(p;p_\text{on}).
\label{Eq:dT0_dLambda_solution}
\end{align}
As was discussed in Sec.~\ref{Sec:nonlocal},
the vertex function $\psi_{\Lambda}$ can be expressed in terms 
of $\psi_{\text{L},\Lambda}$ obtained from the LO potential without a
contact term via
\begin{align}
 \psi_{\Lambda}(p;p_\text{on})=\frac{\psi_{\text{L},\Lambda}(p;p_\text{on})}{1-C_0(\Lambda)\Sigma_{\text{L},\Lambda}(p_\text{on})},
 \label{Eq:psi_Lambda_psi_Lambda_L}
\end{align}
where 
\begin{align}
\Sigma_{\text{L},\Lambda}(p_\text{on})&=\int \frac{p^2 dp}{(2\pi)^3}G_\Lambda(p;p_\text{on})\psi_{\text{L},\Lambda}(p;p_\text{on}).
\label{Eq:Sigma_Lambda_L}
 \end{align}
It is straightforward to verify numerically that both $\psi_{\text{L},\Lambda}$ 
and $\Sigma_{\text{L},\Lambda}$ are natural  (i.e., neither zero nor infinitely
large) at $\Lambda=\bar\Lambda $.
Therefore,
\begin{align}
 \psi_{\Lambda}(p;p_\text{on})\sim \frac{1}{C_0(\Lambda)},\qquad \text{for } \Lambda\sim\ \bar\Lambda ,
\end{align}
so that $\frac{d}{d\Lambda}T^{\rm \scriptscriptstyle LO}$ is finite at
$\Lambda\to\bar\Lambda $. Thus,  
for $\Lambda\sim\bar\Lambda $, one has
\begin{align}
  \frac{d}{d\Lambda}T^{\rm \scriptscriptstyle LO}(p',p;p_\text{on})\sim
  \frac{\psi_{\text{L},\Lambda}(p';p_\text{on})
  \psi_{\text{L},\Lambda}(p;p_\text{on})}
  {\Sigma_{\text{L},\Lambda}(p_\text{on})}
\frac{p_\text{on}^2}{\Lambda^2}.
\label{Eq:limit_dT_dLambda}
\end{align}
As one can see, neglecting the relevant terms in Eq.~\eqref{Eq:dT0_dLambda_2}
altogether is indeed justified\footnote{One still needs to check that
the right-hand side of Eq.~\eqref{Eq:limit_dT_dLambda} tends to zero
as $\Lambda\to\infty$, which can be easily done numerically.}, albeit
this is not the case for neglecting each of them separately.

\section{Renormalizability proof for the model of Long and van Kolck:
  NLO analysis}
\label{Sec:AppenNLO}

We now turn to proving the cutoff independence of the NLO amplitude.
The authors of Ref.~\cite{Long:2007vp} begin their analysis
by taking the derivative $\frac{d}{d\Lambda}T^{\rm \scriptscriptstyle
  NLO}$ and neglecting the terms 
\begin{align}
 2(\mathds{1}+T^{\rm \scriptscriptstyle LO}G_\Lambda)V^{\rm \scriptscriptstyle NLO}G_\Lambda\frac{d T^{\rm \scriptscriptstyle LO}}{d\Lambda}.
\end{align}
This procedure is, however, not justified because the integrals involved in the NLO amplitude
generate positive powers of $\Lambda$, which compensate the negative powers
of $\Lambda$ stemming from $\frac{d T^{\rm \scriptscriptstyle LO}}{d\Lambda}$.
Therefore, we start with separating out the short-range contributions proportional
to $\Lambda$ and $\Lambda^2$ and some other redundant short-range terms and
with expressing the long-range parts of the NLO amplitude in terms of
$T^{\rm \scriptscriptstyle LO}$, for which the $\Lambda$ dependence is already known.

We first notice that the short-range part of the NLO amplitude can be 
rewritten in a simple form using Eq.~\eqref{Eq:psi_prime_via_psi} as
\begin{align}
\label{T_S^NLO}  
T_{\text{S}}^{\rm \scriptscriptstyle NLO}(p_\text{on}) &=
 C_0^{\rm \scriptscriptstyle NLO}(\Lambda)\psi_\Lambda(p_\text{on})^2 
+ 2 C_2^{\rm \scriptscriptstyle NLO}(\Lambda) \psi_\Lambda(p_\text{on}) \psi'_\Lambda(p_\text{on})\nonumber\\
&=\Big[\tilde C_0^{\rm \scriptscriptstyle NLO}(\Lambda)
+\tilde C_2^{\rm \scriptscriptstyle NLO}(\Lambda)p_\text{on}^2\Big]\psi_\Lambda(p_\text{on})^2,
\end{align}
where we have introduced the LECs
\begin{align}
 \tilde C_0^{\rm \scriptscriptstyle NLO}(\Lambda)&=C_0^{\rm \scriptscriptstyle NLO}(\Lambda)
 +2\gamma_1(\Lambda)\Lambda^2 C_2^{\rm \scriptscriptstyle NLO}(\Lambda),\nonumber\\
 \tilde C_2^{\rm \scriptscriptstyle NLO}(\Lambda)&=2\gamma_2 C_2^{\rm \scriptscriptstyle NLO}(\Lambda).
 \label{Eq:C0_C2_tilde}
\end{align}

The long-range parts of the NLO amplitude corresponding to various
terms $V_{\text{L},i}^{\rm \scriptscriptstyle NLO}$
are defined as follows:
\begin{align}
 T_{\text{L},i}^{\rm \scriptscriptstyle NLO}=
 (\mathds{1}+T^{\rm \scriptscriptstyle LO}G_\Lambda) V_{\text{L},i}^{\rm \scriptscriptstyle NLO}
 (\mathds{1}+G_\Lambda T^{\rm \scriptscriptstyle LO}) ,\qquad i=0,1,2.
 \label{Eq:T_L_i}
\end{align}
Using the relation
\begin{align}
 V_{\text{L},0}^{\rm \scriptscriptstyle NLO}
 =-\frac{M_\pi^2}{m_N^2\lambda}V_{\text{L}}^{\rm \scriptscriptstyle LO},
\end{align}
along with the identity
\begin{align}
 (\mathds{1}+T^{\rm \scriptscriptstyle LO}G_\Lambda) V^{\rm \scriptscriptstyle LO}
 (\mathds{1}+G_\Lambda T^{\rm \scriptscriptstyle LO}) 
 =T^{\rm \scriptscriptstyle LO}+
 T^{\rm \scriptscriptstyle LO}G_\Lambda T^{\rm \scriptscriptstyle LO},
\end{align}
one can easily derive the following representation for $T_{\text{L},0}^{\rm \scriptscriptstyle NLO}$: 
\begin{equation}
  \label{T_L0^NLO}
  T_{\text{L},0}^{\rm \scriptscriptstyle NLO}(p_\text{on})=
  \frac{M_\pi^2}{m_N^2\lambda} C_0(\Lambda)\psi_\Lambda(p_\text{on})^2
  - \frac{M_\pi^2}{m_N^2\lambda} T^{\rm \scriptscriptstyle LO}(p_\text{on})
 - \frac{M_\pi^2}{m_N^2\lambda}
 \Big[T^{\rm \scriptscriptstyle LO}G_\Lambda T^{\rm \scriptscriptstyle LO}\Big](p_\text{on}).
\end{equation}

To calculate the amplitude  $T_{\text{L},1}^{\rm \scriptscriptstyle NLO}$,
it is convenient to split the potential $V_{\text{L},1}^{\rm \scriptscriptstyle NLO}$
into three parts via
\begin{align}
  V_{\text{L},1}^{\rm \scriptscriptstyle NLO}&=
  \hat V_{\text{L},1}^{\rm \scriptscriptstyle NLO}
  +V_{\text{L},1,>}^{\rm \scriptscriptstyle NLO}+V_{\text{L},1,<}^{\rm \scriptscriptstyle NLO},\nonumber\\
  \hat V_{\text{L},1}^{\rm \scriptscriptstyle NLO}(p', p;p_\text{on})&=\frac{8\pi^3 }{m_N^3} \frac{p_\text{on}^2}{p_>},\nonumber\\
  V_{\text{L},1,>}^{\rm \scriptscriptstyle NLO}(p', p;p_\text{on})&=
  \frac{8\pi^3 }{m_N^3} \theta(p'-p)\frac{p'^2-p_\text{on}^2}{p'},\nonumber\\
  V_{\text{L},1,<}^{\rm \scriptscriptstyle NLO}(p', p;p_\text{on})&=
  \frac{8\pi^3 }{m_N^3}\theta(p-p')\frac{p^2-p_\text{on}^2}{p},
  \label{Eq:VL1_split}
\end{align}
The quantities $\hat T_{\text{L},1}^{\rm \scriptscriptstyle NLO}$, $T_{\text{L},1,>}^{\rm \scriptscriptstyle NLO}$
and $T_{\text{L},1,<}^{\rm \scriptscriptstyle NLO}$ are defined analogously to Eq.~\eqref{Eq:T_L_i}.
The amplitude $\hat T_{\text{L},1}^{\rm \scriptscriptstyle NLO}$ can be calculated 
similarly to $T_{\text{L},0}^{\rm \scriptscriptstyle NLO}$:
\begin{equation}
  \label{Eq:HatT_L1^NLO}
  \hat T_{\text{L},1}^{\rm \scriptscriptstyle NLO}(p_\text{on}) =
  \frac{p_\text{on}^2}{m_N^2\lambda} C_0(\Lambda)\psi_\Lambda(p_\text{on})^2
  - \frac{p_\text{on}^2}{m_N^2\lambda} T^{\rm \scriptscriptstyle LO}(p_\text{on})
 - \frac{p_\text{on}^2}{m_N^2\lambda}
 \Big[T^{\rm \scriptscriptstyle LO}G_\Lambda T^{\rm \scriptscriptstyle LO}\Big](p_\text{on}).
\end{equation}
To calculate $T_{\text{L},1,<}^{\rm \scriptscriptstyle NLO}$ on shell,
we perform a single iteration of 
the LO Lippmann-Schwinger equation:
\begin{align}
(\mathds{1}+T^{\rm \scriptscriptstyle LO}G_\Lambda) V_{\text{L},1,<}^{\rm \scriptscriptstyle NLO}
 (\mathds{1}+G_\Lambda T^{\rm \scriptscriptstyle LO})  = 
 (\mathds{1}+T^{\rm \scriptscriptstyle LO}G_\Lambda) V_{\text{L},1,<}^{\rm \scriptscriptstyle NLO}
 G_\Lambda V^{\rm \scriptscriptstyle LO}
 (\mathds{1}+G_\Lambda T^{\rm \scriptscriptstyle LO}).
\end{align}
The explicit evaluation of $V_{\text{L},1,<}^{\rm \scriptscriptstyle NLO}
 G_\Lambda V^{\rm \scriptscriptstyle LO}$ yields
\begin{align}
\Big[V_{\text{L},1,<}^{\rm \scriptscriptstyle NLO}
 G_\Lambda V^{\rm \scriptscriptstyle LO}\Big](p',p;p_\text{on})&=
 \int \frac{p''^2dp''}{(2\pi)^3}
 \frac{8\pi^3 }{m_N^3} \theta(p''-p')\frac{p''^2-p_\text{on}^2}{p''}
 G_\Lambda(p'';p_\text{on})
 V^{\rm \scriptscriptstyle LO}(p'',p)\nonumber\\
 &=-\frac{1}{m_N^2}\int_{p'}^{\Lambda} p''dp''
 \Big[C_0(\Lambda)-\frac{8\pi^3\lambda}{m_N p''}\theta(p''-p)-\frac{8\pi^3\lambda}{m_N p}\theta(p-p'')\Big]\nonumber\\
 &=\frac{C_0(\Lambda)}{2m_N^2}(p'^2-\Lambda^2) 
 +\frac{8\pi^3\lambda}{m_N^3}\Lambda
 -\frac{8\pi^3\lambda}{2 m_N^3 p}\theta(p-p')(p^2+p'^2)
 -\frac{8\pi^3\lambda}{m_N^3}\theta(p'-p)p'.
  \end{align}
Symmetrically, to calculate $T_{\text{L},1,>}^{\rm \scriptscriptstyle NLO}$ on shell, we iterate
the LO Lippmann-Schwinger equation on the left,
\begin{align}
(\mathds{1}+T^{\rm \scriptscriptstyle LO}G_\Lambda) V_{\text{L},1,<}^{\rm \scriptscriptstyle NLO}
 (\mathds{1}+G_\Lambda T^{\rm \scriptscriptstyle LO})  = 
 (\mathds{1}+T^{\rm \scriptscriptstyle LO}G_\Lambda)
  V^{\rm \scriptscriptstyle LO}G_\Lambda
 V_{\text{L},1,<}^{\rm \scriptscriptstyle NLO}
  (\mathds{1}+G_\Lambda T^{\rm \scriptscriptstyle LO}),
\end{align}
and obtain
\begin{align}
\Big[V^{\rm \scriptscriptstyle LO}G_\Lambda V_{\text{L},1,>}^{\rm \scriptscriptstyle NLO}
 \Big](p',p;p_\text{on})
 &=\frac{C_0(\Lambda)}{2m_N^2}(p^2-\Lambda^2) 
 +\frac{8\pi^3\lambda}{m_N^3}\Lambda
 -\frac{8\pi^3\lambda}{2 m_N^3 p'}\theta(p'-p)(p^2+p'^2)
 -\frac{8\pi^3\lambda}{m_N^3}\theta(p-p')p.
  \end{align}
Combining the two pieces yields
\begin{align}
\Big[V_{\text{L},1,<}^{\rm \scriptscriptstyle NLO}
 G_\Lambda V^{\rm \scriptscriptstyle LO}
 +V^{\rm \scriptscriptstyle LO}G_\Lambda V_{\text{L},1,>}^{\rm \scriptscriptstyle NLO}
 \Big](p',p;p_\text{on})
 &=-\frac{\mathcal{\tilde M}(\Lambda)}{m_N^2}\Lambda^2
+\frac{C_0(\Lambda)}{2m_N^2}(p^2+p'^2)
 -\frac{8\pi^3\lambda}{2 m_N^3 p_>}(3p_>^2+p_<^2),
\end{align}
with
\begin{align}
\mathcal{\tilde M}(\Lambda)=
 -2\frac{(2\pi)^3\lambda}{m_N\Lambda}+C_0(\Lambda),
 \label{Eq:defifition_Mcal_tilde}
\end{align}
and we finally obtain for the $T_{\text{L},1}^{\rm \scriptscriptstyle NLO}$ amplitude:
\begin{align}
 T_{\text{L},1}^{\rm \scriptscriptstyle NLO}(p_\text{on})
 &=\hat T_{\text{L},1}^{\rm \scriptscriptstyle NLO}(p_\text{on})
 -\frac{\mathcal{\tilde M}(\Lambda)}{m_N^2}\Lambda^2\psi_\Lambda(p_\text{on})^2
 +\frac{C_0(\Lambda)}{m_N^2}\psi_\Lambda(p_\text{on})\psi'_\Lambda(p_\text{on})
 -\frac{3\lambda}{2}T_{\text{L},1}^{\rm \scriptscriptstyle NLO}(p_\text{on})
 -\frac{\lambda}{2}T_{\text{L},2}^{\rm \scriptscriptstyle NLO}(p_\text{on}).
 \label{Eq:TL1_NLO}
\end{align}

We treat $V_{\text{L},2}^{\rm \scriptscriptstyle NLO}$ analogously 
to $V_{\text{L},1}^{\rm \scriptscriptstyle NLO}$ in Eq.~\eqref{Eq:VL1_split} and split it as follows:
\begin{align}
  V_{\text{L},2}^{\rm \scriptscriptstyle NLO}&=
  \hat V_{\text{L},2}^{\rm \scriptscriptstyle NLO}
  +V_{\text{L},2,>}^{\rm \scriptscriptstyle NLO}+V_{\text{L},2,<}^{\rm \scriptscriptstyle NLO},\nonumber\\
  \hat V_{\text{L},2}^{\rm \scriptscriptstyle NLO}(p', p;p_\text{on})&=\frac{8\pi^3 }{m_N^3} \frac{p_\text{on}^2}{p_>},\nonumber\\
  V_{\text{L},2,>}^{\rm \scriptscriptstyle NLO}(p', p;p_\text{on})&=
  \frac{8\pi^3 }{m_N^3} \theta(p'-p)\frac{p^2-p_\text{on}^2}{p'},\nonumber\\
  V_{\text{L},2,<}^{\rm \scriptscriptstyle NLO}(p', p;p_\text{on})&=
  \frac{8\pi^3 }{m_N^3}\theta(p-p')\frac{p'^2-p_\text{on}^2}{p}.
\end{align}
Performing the same manipulations with the amplitude $T_{\text{L},2}^{\rm \scriptscriptstyle NLO}$ as above, we obtain:
\begin{align}
 \hat T_{\text{L},2}^{\rm \scriptscriptstyle NLO}(p_\text{on})&=\hat T_{\text{L},1}^{\rm \scriptscriptstyle NLO}(p_\text{on}),\nonumber\\
 T_{\text{L},2}^{\rm \scriptscriptstyle NLO}(p_\text{on})&=
 \hat T_{\text{L},1}^{\rm \scriptscriptstyle NLO}(p_\text{on})
- \frac{2 C_0(\Lambda)}{3 m_N^2}\psi_\Lambda(p_\text{on})\psi'_\Lambda(p_\text{on})
 +\frac{\lambda}{2}T_{\text{L},1}^{\rm \scriptscriptstyle NLO}(p_\text{on})
 +\frac{\lambda}{6}T_{\text{L},2}^{\rm \scriptscriptstyle NLO}(p_\text{on}).
\label{Eq:TL2_NLO} 
\end{align}
To derive the latter equation,
we used the result of the integral
\begin{align}
\Big[V_{\text{L},2,>}^{\rm \scriptscriptstyle NLO}
 G_\Lambda V^{\rm \scriptscriptstyle LO}\Big](p',p;p_\text{on})&=
 \int \frac{p''^2dp''}{(2\pi)^3}
 \frac{8\pi^3 }{m_N^3} \theta(p'-p'')\frac{p''^2-p_\text{on}^2}{p'}
 G_\Lambda(p'';p_\text{on})
 V^{\rm \scriptscriptstyle LO}(p'',p)\nonumber\\
 &=-\frac{1}{m_N^2 p'}\int_0^{p'} p''^2dp''
 \Big[C_0(\Lambda)-\frac{8\pi^3\lambda}{m_N p''}\theta(p''-p)-\frac{8\pi^3\lambda}{m_N p}\theta(p-p'')\Big]\nonumber\\
 &=-\frac{C_0(\Lambda)}{3 m_N^2}p'^2
 +\frac{8\pi^3\lambda}{2 m_N^3 p'}\theta(p'-p)\Big(p'^2-\frac{p^2}{3}\Big)
 +\frac{8\pi^3\lambda}{3 m_N^3 p}\theta(p-p')p'^2,
  \end{align}
  and the expression for the symmetric combination
\begin{align}
\Big[V_{\text{L},2,>}^{\rm \scriptscriptstyle NLO}
 G_\Lambda V^{\rm \scriptscriptstyle LO}
 +V^{\rm \scriptscriptstyle LO}G_\Lambda V_{\text{L},2,<}^{\rm \scriptscriptstyle NLO}
 \Big](p',p;p_\text{on})
 =-\frac{C_0(\Lambda)}{3 m_N^2}(p^2+p'^2)
 +\frac{8\pi^3\lambda}{2 m_N^3 p_>}\Big(p_>^2+\frac{p_<^2}{3}\Big).
\end{align}
Finally, solving the system of equations~\eqref{Eq:TL1_NLO}~and~\eqref{Eq:TL2_NLO}, we obtain
\begin{align}
 T_{\text{L},1}^{\rm \scriptscriptstyle NLO}(p_\text{on})&=
 \frac{1}{2(3+4\lambda)}
 \bigg[(6-4\lambda)\hat T_{\text{L},1}^{\rm \scriptscriptstyle NLO}(p_\text{on})
 +(\lambda-6)\frac{\mathcal{\tilde M}(\Lambda)}{m_N^2}\Lambda^2\psi_\Lambda(p_\text{on})^2
+(6+\lambda)\frac{C_0(\Lambda)}{m_N^2}\psi_\Lambda(p_\text{on})\psi'_\Lambda(p_\text{on})
 \bigg],
  \label{Eq:T_L1^NLO_final} \\
 T_{\text{L},2}^{\rm \scriptscriptstyle NLO}(p_\text{on})&=
 \frac{1}{2(3+4\lambda)}
 \bigg[(6+12\lambda)\hat T_{\text{L},1}^{\rm \scriptscriptstyle NLO}(p_\text{on})
 -3\lambda\frac{\mathcal{\tilde M}(\Lambda)}{m_N^2}\Lambda^2\psi_\Lambda(p_\text{on})^2
 -(4+3\lambda)\frac{C_0(\Lambda)}{m_N^2}\psi_\Lambda(p_\text{on})\psi'_\Lambda(p_\text{on})
 \bigg].
 \label{Eq:T_L2^NLO_final} 
\end{align}

We now combine everything together. Using the expressions for the
amplitudes $T_{\text{S}}^{\rm \scriptscriptstyle NLO}(p_\text{on})$,
$T_{\text{L},0}^{\rm \scriptscriptstyle NLO}(p_\text{on})$,
$T_{\text{L},1}^{\rm \scriptscriptstyle NLO}(p_\text{on})$ and
$T_{\text{L},2}^{\rm \scriptscriptstyle NLO}(p_\text{on})$ given in 
Eqs.~\eqref{T_S^NLO}, \eqref{T_L0^NLO}, \eqref{Eq:T_L1^NLO_final} and \eqref{Eq:T_L2^NLO_final}, 
respectively, and taking into account
Eqs.~\eqref{Eq:HatT_L1^NLO}~and~\eqref{Eq:psi_prime_via_psi}, 
the full NLO amplitude can be represented as the sum
\begin{align}
 T^{\rm \scriptscriptstyle NLO}(p_\text{on})
 =T_0^{\rm \scriptscriptstyle NLO}(p_\text{on})+T_2^{\rm \scriptscriptstyle NLO}(p_\text{on}) ,
\end{align}
with
\begin{eqnarray}
 T_0^{\rm \scriptscriptstyle NLO}(p_\text{on})
 &=&- g_0\frac{M_\pi^2}{m_N^2\lambda} 
 \Big\{T^{\rm \scriptscriptstyle LO}(p_\text{on})
 +
 \Big[T^{\rm \scriptscriptstyle LO}G_\Lambda T^{\rm \scriptscriptstyle LO}\Big](p_\text{on})\Big\}
 +\bar C_0^{\rm \scriptscriptstyle
  NLO}(\Lambda)\psi_\Lambda(p_\text{on})^2, \\
 T_2^{\rm \scriptscriptstyle NLO}(p_\text{on})
 &=&- \bar g\frac{p_\text{on}^2}{m_N^2\lambda} 
 \Big\{T^{\rm \scriptscriptstyle LO}(p_\text{on})
 +
 \Big[T^{\rm \scriptscriptstyle LO}G_\Lambda T^{\rm \scriptscriptstyle LO}\Big](p_\text{on})\Big\}
 +\bar C_2^{\rm \scriptscriptstyle NLO}(\Lambda)\psi_\Lambda(p_\text{on})^2 p_\text{on}^2,
\end{eqnarray}
where we have introduced the constant
\begin{equation}
\bar g =\frac{(3-2\lambda)g_1+(3+6\lambda)g_2}{3+4\lambda}.
\end{equation}
The new LECs $\bar C_0^{\rm \scriptscriptstyle NLO}(\Lambda)$
and $\bar C_2^{\rm \scriptscriptstyle NLO}(\Lambda)$
absorb the redundant short-range contributions and are given by
\begin{align}
\bar C_0^{\rm \scriptscriptstyle NLO}(\Lambda)&=\tilde C_0^{\rm \scriptscriptstyle NLO}(\Lambda)
 +\frac{g_0}{\lambda}\frac{C_0(\Lambda)}{m_N^2}M_\pi^2
 +\gamma_1(\Lambda) \frac{(6+\lambda)g_1-(4+3\lambda) g_2}{2(3+4\lambda)}\frac{C_0(\Lambda)}{m_N^2}\Lambda^2  +\frac{(\lambda-6)g_1-3\lambda g_2}{2(3+4\lambda)}\frac{\mathcal{\tilde M}(\Lambda)}{m_N^2}\Lambda^2
 ,\nonumber\\
 \bar C_2^{\rm \scriptscriptstyle NLO}(\Lambda)&=\tilde C_2^{\rm \scriptscriptstyle NLO}(\Lambda)
 + \Big[\frac{\bar g}{\lambda}+\gamma_2  \frac{(6+\lambda)g_1-(4+3\lambda) g_2}{2(3+4\lambda)}\Big]\frac{C_0(\Lambda)}{m_N^2}.
 \label{Eq:C0_C2_bar}
\end{align}
Now that all positive powers of the cutoff are absorbed by the above redefinition of 
the NLO LECs, we are indeed in the position to derive the equations
the constants $\bar C_0^{\rm \scriptscriptstyle NLO}(\Lambda)$
and $\bar C_2^{\rm \scriptscriptstyle NLO}(\Lambda)$ have to fulfill
to  make the NLO amplitude cutoff independent.
Obviously, both $T_0^{\rm \scriptscriptstyle NLO}$ and $T_2^{\rm \scriptscriptstyle NLO}$
can be made cutoff independent individually, and the equations for $\bar C_0^{\rm \scriptscriptstyle NLO}(\Lambda)$
and $\bar C_2^{\rm \scriptscriptstyle NLO}(\Lambda)$ decouple.

We start with the amplitude  $dT_0^{\rm \scriptscriptstyle NLO}(p_\text{on})/d\Lambda$:
\begin{align}
 \frac{d}{d\Lambda}T_0^{\rm \scriptscriptstyle NLO}(p_\text{on})&=
 - g_0\frac{M_\pi^2}{m_N^2\lambda} 
 \Big\{\frac{d}{d\Lambda}T^{\rm \scriptscriptstyle LO}(p_\text{on})
 +
 \Big[T^{\rm \scriptscriptstyle LO}\frac{dG_\Lambda}{d\Lambda} T^{\rm \scriptscriptstyle LO}\Big](p_\text{on})
 +2\Big[T^{\rm \scriptscriptstyle LO}G_\Lambda \frac{d}{d\Lambda}T^{\rm \scriptscriptstyle LO}\Big](p_\text{on})\Big\}
 \nonumber\\
 &+\psi_\Lambda(p_\text{on})^2\frac{d}{d\Lambda}\bar C_0^{\rm \scriptscriptstyle NLO}(\Lambda)
 +2\bar C_0^{\rm \scriptscriptstyle NLO}(\Lambda)\psi_\Lambda(p_\text{on})\frac{d}{d\Lambda}\psi_\Lambda(p_\text{on}).
 \label{Eq:dT_dLambda_start}
\end{align}
Using Eqs.~\eqref{Eq:T_Lambda} and \eqref{Eq:dT0_dLambda_solution},
along with the definition~\eqref{Eq:psi_psi_prime}, we obtain for the 
derivative $\frac{d}{d\Lambda}\psi_\Lambda(p_\text{on})$:
\begin{align}
 \frac{d}{d\Lambda}\psi_\Lambda(p_\text{on})=
 -\frac{\mathcal{M}(\Lambda)m_N}{(2\pi)^3}\psi_\Lambda(p_\text{on})
 \Big\{1
 +\frac{p_\text{on}^2}{\Lambda^2-p_\text{on}^2}
 \Big[1+\mathcal{M}(\Lambda)\Sigma_\Lambda(p_\text{on})\Big]
\Big\},
\end{align}
where 
\begin{align}
\Sigma_{\Lambda}(p_\text{on})&=\int \frac{p^2 dp}{(2\pi)^3}G_\Lambda(p;p_\text{on})\psi_{\Lambda}(p;p_\text{on})
=\frac{\Sigma_{\text{L},\Lambda}(p_\text{on})}{1-C_0(\Lambda)\Sigma_{\text{L},\Lambda}(p_\text{on})}.
\label{Eq:Sigma_SigmaL}
 \end{align}
Performing the same manipulations with the other terms in Eq.~\eqref{Eq:dT_dLambda_start},
we find
\begin{align}
 \frac{d}{d\Lambda}T_0^{\rm \scriptscriptstyle NLO}(p_\text{on})&=
 \bigg[\frac{d}{d\Lambda}\bar C_0^{\rm \scriptscriptstyle NLO}(\Lambda)
 -2\bar C_0^{\rm \scriptscriptstyle NLO}(\Lambda)
 \frac{\mathcal{M}(\Lambda)m_N}{(2\pi)^3}
 +g_0\frac{M_\pi^2}{m_N\lambda} \frac{\mathcal{M}(\Lambda)^2}{(2\pi)^3}
 \bigg]\psi_\Lambda(p_\text{on})^2
 \nonumber\\
 &+\frac{p_\text{on}^2}{\Lambda^2-p_\text{on}^2}
 \bigg\{
 2g_0\frac{M_\pi^2}{m_N\lambda} \frac{\mathcal{M}(\Lambda)^2}{(2\pi)^3}
 \psi_\Lambda(p_\text{on})\big[\tilde\Sigma_\Lambda(p_\text{on})+\psi_\Lambda(p_\text{on})\big]
   \nonumber\\
 & -2\bar C_0^{\rm \scriptscriptstyle NLO}(\Lambda)
 \frac{\mathcal{M}(\Lambda)m_N}{(2\pi)^3}
 \big[1+\mathcal{M}(\Lambda)\Sigma_\Lambda(p_\text{on})\big]\psi_\Lambda(p_\text{on})^2
\bigg\}.
 \label{Eq:dT_dLambda_2}
\end{align}
where
\begin{align}
\tilde\Sigma_{\Lambda}(p_\text{on})&=\int \frac{p^2 dp}{(2\pi)^3}
T_0^{\rm \scriptscriptstyle NLO}(p_\text{on},p;p_\text{on})G_\Lambda(p;p_\text{on})\psi_{\Lambda}(p;p_\text{on}).
\label{Eq:Sigma_tilde}
 \end{align}
Next, we assume that the order $O(1/\Lambda^2)$ terms can be neglected (to be justified afterwards)
and obtain the following equation for $\bar C_0^{\rm \scriptscriptstyle NLO}(\Lambda)$:
\begin{align}
 \frac{d}{d\Lambda}\bar C_0^{\rm \scriptscriptstyle NLO}(\Lambda)
 -2\bar C_0^{\rm \scriptscriptstyle NLO}(\Lambda)
 \frac{\mathcal{M}(\Lambda)m_N}{(2\pi)^3}
 +g_0\frac{M_\pi^2}{m_N\lambda} \frac{\mathcal{M}(\Lambda)^2}{(2\pi)^3}=0.
 \label{Eq:equation_C_0_NLO}
\end{align}
We look for a solution $H(\Lambda)$  of the corresponding homogeneous
differential equation by
explicitly factorizing out the singularity close to $\Lambda=\bar\Lambda$.
This can be conveniently done by defining
\begin{align}
 H(\Lambda)\eqqcolon
 \big[\mathcal{M}(\Lambda)^2+\Delta(\Lambda)\big]\tilde H(\Lambda),
 \end{align}
with
 \begin{align}
 \Delta(\Lambda)=\frac{(2\pi)^6\lambda}{m_N^2\Lambda^2}.
\end{align}
Then, using Eq.~\eqref{Eq:equation_for_C0},
we obtain the equation for $\tilde H(\Lambda)$,
\begin{align}
 \frac{d}{d\Lambda}\tilde H(\Lambda)=2\frac{(2\pi)^6\lambda}{m_N^2\Lambda^3}
 \big[\mathcal{M}(\Lambda)^2+\Delta(\Lambda)\big]^{-1}\tilde H(\Lambda),
\end{align}
which has a solution 
\begin{align}
 \tilde H(\Lambda)=
  \exp\bigg\{
  {2\frac{(2\pi)^6\lambda}{m_N^2}\int_{\Lambda_*}^\Lambda\frac{d\Lambda}{\Lambda^3[\mathcal{M}(\Lambda)^2+\Delta(\Lambda)]}}
  \bigg\},
  \end{align}
where we have used for the normalization the same quantity $\Lambda_*$ as in $C_0(\Lambda)$.
The solution of the inhomogeneous equation~\eqref{Eq:equation_C_0_NLO} is readily obtained:
\begin{align}
  \label{Eq:LambdaBarC_0^NLO}
 \bar C_0^{\rm \scriptscriptstyle NLO}(\Lambda)=H(\Lambda)
 \bigg[\frac{\bar C_0^{\rm \scriptscriptstyle NLO}(\Lambda_*)}{H(\Lambda_*)}
 -\frac{g_0 M_\pi^2}{(2\pi)^3 m_N\lambda}
 \int_{\Lambda_*}^\Lambda\frac{d\Lambda \mathcal{M}(\Lambda)^2}{H(\Lambda)} \bigg].
\end{align}
Analogously to the case of $\bar C_0^{\rm \scriptscriptstyle NLO}(\Lambda)$,
we
derive an equation and a solution for $\bar C_2^{\rm \scriptscriptstyle NLO}(\Lambda)$:
\begin{align}
 \frac{d}{d\Lambda}\bar C_2^{\rm \scriptscriptstyle NLO}(\Lambda)
 -2\bar C_2^{\rm \scriptscriptstyle NLO}(\Lambda)
 \frac{\mathcal{M}(\Lambda)m_N}{(2\pi)^3}
 +\bar g\frac{1}{m_N\lambda} \frac{\mathcal{M}(\Lambda)^2}{(2\pi)^3}=0,
 \label{Eq:equation_C_2_NLO}
\end{align}
\begin{align}
 \bar C_2^{\rm \scriptscriptstyle NLO}(\Lambda)=H(\Lambda)
 \bigg[\frac{\bar C_2^{\rm \scriptscriptstyle NLO}(\Lambda_*)}{H(\Lambda_*)}
 -\frac{\bar g}{(2\pi)^3 m_N\lambda}
 \int_{\Lambda_*}^\Lambda\frac{d\Lambda \mathcal{M}(\Lambda)^2}{H(\Lambda)} \bigg].
\label{Eq:LambdaBarC_2^NLO}
 \end{align}
Notice that $H(\Lambda)\sim \mathcal{M}(\Lambda)^2\sim C_0(\Lambda)^2$ for 
cutoff values $\Lambda\sim \bar\Lambda$.
Therefore, as expected, the positions of the singularities of $\bar C_0^{\rm \scriptscriptstyle NLO}(\Lambda)$
and $\bar C_2^{\rm \scriptscriptstyle NLO}(\Lambda)$ coincide with the
ones of $C_0(\Lambda)$, i.e., $\Lambda=\bar\Lambda$,
and the behavior of $\bar C_0^{\rm \scriptscriptstyle NLO}(\Lambda)$
and $\bar C_2^{\rm \scriptscriptstyle NLO}(\Lambda)$ in the vicinity of $\bar\Lambda$ is given by
\begin{align}
 \bar C_0^{\rm \scriptscriptstyle NLO}(\Lambda)\sim\bar C_2^{\rm \scriptscriptstyle NLO}(\Lambda)
 \sim(\Lambda-\bar\Lambda)^2.
 \label{Eq:singularity_C_0_C_2}
\end{align}

The last step in proving the cutoff independence of the NLO amplitude is 
to show that the neglected terms of order $\sim p_\text{on}^2/\Lambda^2$
are suppressed also in the vicinity of $\Lambda=\bar\Lambda$ \footnote{For regular cutoffs, 
this feature can be straightforwardly verified numerically.}.
Inspecting Eq.~\eqref{Eq:dT_dLambda_2} we see that 
the terms proportional to $\mathcal{M}(\Lambda)^2)\psi_\Lambda(p_\text{on})^2$
and $\mathcal{M}(\Lambda)^2\psi_\Lambda(p_\text{on})\tilde\Sigma(\Lambda)$
are regular at $\Lambda=\bar\Lambda $ as follows from Eq.~\eqref{Eq:psi_Lambda_psi_Lambda_L}
and the definition of $\tilde\Sigma$ in Eq.~\eqref{Eq:Sigma_tilde}.
The remaining term proportional to 
$\bar C_0^{\rm \scriptscriptstyle NLO}(\Lambda)\mathcal{M}(\Lambda)\big[1+\mathcal{M}(\Lambda)\Sigma_\Lambda(p_\text{on})\big]\psi_\Lambda(p_\text{on})^2$
can be rewritten as (see Eqs.~\eqref{Eq:Sigma_SigmaL}~and~\eqref{Eq:psi_Lambda_psi_Lambda_L}):
\begin{align}
 \bar C_0^{\rm \scriptscriptstyle NLO}(\Lambda)\mathcal{M}(\Lambda)
 \frac{m_N}{(2\pi)^3}
 \big[1+\mathcal{M}(\Lambda)\Sigma_\Lambda(p_\text{on})\big]\psi_\Lambda(p_\text{on})^2=
 \left[\frac{m_N}{(2\pi)^3}-\lambda\frac{\Sigma_{\text{L},\Lambda}(p_\text{on})}{\Lambda}\right]
  \frac{\bar C_0^{\rm \scriptscriptstyle NLO}(\Lambda)\mathcal{M}(\Lambda)\psi_{\text{L},\Lambda}(p_\text{on})^2}
 {\big[1-C_0(\Lambda)\Sigma_{\text{L},\Lambda}(p_\text{on})\big]^3},
\end{align}
and is, therefore, also regular at $\Lambda=\bar\Lambda $ given the 
behaviour of $\bar C_0^{\rm \scriptscriptstyle NLO}(\Lambda)$
in Eq.~\eqref{Eq:singularity_C_0_C_2}.

The original constants $C_0^{\rm \scriptscriptstyle NLO}(\Lambda)$
and $C_2^{\rm \scriptscriptstyle NLO}(\Lambda)$
can be reconstructed from the solutions for $\bar C_0^{\rm \scriptscriptstyle NLO}(\Lambda)$
and $\bar C_2^{\rm \scriptscriptstyle NLO}(\Lambda)$
using Eqs.~\eqref{Eq:C0_C2_tilde}~and~\eqref{Eq:C0_C2_bar}.
We have verified numerically that the analytical 
solutions for $C_0^{\rm \scriptscriptstyle NLO}(\Lambda)$
and $C_2^{\rm \scriptscriptstyle NLO}(\Lambda)$ indeed lead to 
the cutoff independent NLO amplitude in the limit $\Lambda\to\infty$.

\bibliography{4.9}
\bibliographystyle{apsrev}

\end{document}